\documentstyle[12pt,ssi]{article}
\input epsf
\newcommand{\beq}{\begin{equation}}
\newcommand{\eeq}{\end{equation}}
\newcommand{\beqa}{\begin{eqnarray}}
\newcommand{\eeqa}{\end{eqnarray}}

\newcommand{\dphi}{\mbox{$\delta \phi$}}
\newcommand{\bbbar}{\mbox{$B-\bar B$}}
\newcommand{\apks}{\mbox{$a_{\psi K_S}$}}
\newcommand{\afks}{\mbox{$a_{\phi K_S}$}}
\newcommand{\app}{\mbox{$a_{\pi\pi}$}}
\newcommand{\epsK}{\mbox{$\epsilon_K$}}
\newcommand{\stb}{\mbox{$\sin2\beta$}}
\newcommand{\kpnn}{\mbox{$K\to\pi\nu\bar\nu$}}
\newcommand{\apnn}{\mbox{$a_{\pi\nu\bar\nu}$}}
\def\UT{unitarity triangle}
\def\sm{Standard Model}
\def\np{new physics}
\def\gsim{{~\raise.15em\hbox{$>$}\kern-.85em
          \lower.35em\hbox{$\sim$}~}}
\def\lsim{{~\raise.15em\hbox{$<$}\kern-.85em
          \lower.35em\hbox{$\sim$}~}}

\def\ie{{\it i.e.}}
\def\eg{{\it e.g.}}
\def\etal{{\it et al}}
\newcommand{\lpp}{\mbox{$\lambda^{\prime \prime}$}} 
\newcommand{\qt}{\mbox{$\tilde q$}}
\def\npb#1{Nucl.\ Phys.\ {\bf B #1}}
\def\plb#1{Phys.\ Lett.\ {\bf B #1}}
\def\prd#1{Phys.\ Rev.\ {\bf D #1}}
\def\prl#1{Phys.\ Rev.\ Lett. {\bf #1}}

\def\zpc#1{Z.~Phys.\ {\bf C #1}}

\def\ijmpa#1{Int.\ J.\ Mod.\ Phys.\ {\bf A #1}}
\def\progtp#1{Prog.\ Th.\ Phys.\ {\bf #1}}

\begin{document}

\rightline{SLAC-PUB-7700}
\rightline{hep-ph/9711265}
\rightline{November 1997}

\title{Searching for New Physics with ${\bf CP}$ Violating 
       ${\bf B}$ Decays}

\author{Mihir P. Worah\thanks{Supported by DOE Contract DE-ACO3-76SF00515. 
\vskip 0.5in 
\noindent
\copyright\ 1997 by Mihir P. Worah. }\\ 
Stanford Linear Accelerator Center \\
Stanford University, Stanford, CA 94309 \\[0.4cm]
}

\maketitle
\begin{abstract}
We explore the possibility of using $CP$ violation in $B$
decays to detect the presence of physics beyond the \sm. 
We first study the possibility of new physics in the $B-\bar B$ mixing
amplitude. We discuss a construction to extract information 
about the phase and
magnitude of the new physics contribution, as well as the CKM
parameters in a model independent way. We point out the difficulty of
carrying through this program induced by hadronic uncertainties and
discrete ambiguities, and suggest additional measurements 
to overcome these problems.  
We then study the possibility of new physics contributions to the $B$
meson decay amplitudes. We emphasize the sensitivity of the $B\to \phi
K_S$ decay to these new contributions, and explain how this
sensitivity can be quantified using experimental data on $SU(3)$
related decays. Finally, we analyse
a number of models where the $B$ decay amplitudes are modified. 
\end{abstract}

\section{Introduction}

$CP$ violation has so far only been observed in the decays of neutral
$K$ mesons. It is one of the goals of the proposed $B$ factories
to find and study $CP$ violation in the decays of $B$
mesons, and thus elucidate the mechanisms by which $CP$
violation manifests itself in the low energy world. 
There is a commonly accepted \sm\ of $CP$ violation,
namely that it is a result of the one physical phase in the $3 \times 3$
Cabbibo Kobayashi Maskawa (CKM) matrix \cite{ckm}. 
This scenario has specific predictions for the magnitude as well as
patterns of $CP$ violation that will be observed in the $B$ meson
decays \cite{Yossi}. 
However, since there currently exists only one experimental measurement of
$CP$ violation, it is possible that the \sm\ explanation for it is
incorrect, or more likely that in addition to the one CKM phase, there
are additional $CP$ violating phases introduced by whatever new
physics lies beyond the \sm.

In this lecture we study the possibility of detecting the presence of
physics beyond the \sm, using the $CP$ violating asymmetries measured in
the decays of neutral $B_d$ mesons to $CP$ eigenstates, in a 
largely model independent way. (For recent reviews concerning possible
outcomes in specific models see Refs. \citenum{gronau-london96,GNR}). 
We first 
introduce the necessary formalism and, in Sec. 2, briefly review
the situation concerning these $CP$ asymmetries in the \sm. Sec. 3 deals
with the possibility of new physics in the \bbbar\ mixing
amplitude, while in Sec. 4 we study the possibility of new physics 
in the $B$ decay amplitudes. We present our conclusions in Sec. 5.
    
\subsection{Formalism}

In this sub-section we display the well known formulae for the decays
of neutral $B$ mesons into $CP$ eigenstates, and highlight the 
relevant features that are important when more than one decay
amplitude contribute to a particular process.

The time dependent $CP$ asymmetry for the decays of states that
were tagged as pure $B^0$ or $\bar B^0$ at production into $CP$
eigenstates is defined as
\beq
a_{f_{CP}}(t) \equiv \frac{\Gamma[B^0(t) \rightarrow f_{CP}] -
                           \Gamma[\bar B^0(t) \rightarrow f_{CP}]}
                           {\Gamma[B^0(t) \rightarrow f_{CP}] +
                           \Gamma[\bar B^0(t) \rightarrow f_{CP}]},
\eeq
and given by
\beq
a_{f_{CP}}(t) = a_{f_{CP}}^{cos}\cos(\Delta Mt) + 
                a_{f_{CP}}^{sin}\sin(\Delta Mt) 
\label{cos-sin}
\eeq
where 
\beq
a_{f_{CP}}^{cos} = \frac{(1-|\lambda|^2)}{1 + |\lambda|^2};~~~~
a_{f_{CP}}^{sin} = -\frac{2 {\rm Im} \lambda}
                    {1 +|\lambda|^2}.
\label{aCP}
\eeq
Here $\Delta M$ is the mass difference between the two physical
states, and 
\beq
\lambda = \left(\sqrt{\frac{M_{12}^{*}-\frac{i}{2}\Gamma_{12}^{*}}
               {M_{12}-\frac{i}{2}\Gamma_{12}}}\right)
          \frac{\langle f_{CP} |{\cal H} | \bar B^0 \rangle}
              {\langle f_{CP} |{\cal H} | B^0 \rangle}
        = e^{-2i\phi_M}\frac{\bar A}{A},  
\label{lambda}
\eeq
where we have used the fact that $M_{12} \gg \Gamma_{12}$, to replace
the first fraction in Eq.~(\ref{lambda}) by $e^{-2i\phi_M}$, the phase
of $B-\bar B$ mixing.

If the decay amplitude $A$ has only one dominant contribution, 
$A = |A|e^{i\phi_D}$, then one has $\bar A = A^*$ and consequently
$|\lambda | =1$. Thus, in this case, $a_{f_{CP}}^{cos} = 0$, and 
$a_{f_{CP}}^{sin} = \sin 2(\phi_M+\phi_D)$ is a clean measure of the $CP$
violation due to interference between the mixing and decay
amplitudes. In addition, if there is no new physics
contribution to the mixing matrix (or if it is in phase with the \sm\
contribution), $a_{f_{CP}}^{sin}$ 
cleanly measures $CP$ violating phases in
the CKM matrix since both $\phi_M$ and $\phi_D$ are simply sums of 
these.\cite{isi-jon}

Consider now the case where the decay
amplitude $A$ contains contributions from two terms with 
magnitudes $A_i$,
$CP$ violating phases $\phi_i$ and $CP$ conserving phases $\delta_i$
(in what follows it will be convenient to think of $A_1$ giving the
dominant \sm\ contribution, and $A_2$ giving the sub leading \sm\
contribution or the new physics contribution).
\beq
A = A_1e^{i\phi_1}e^{i\delta_1} + A_2e^{i\phi_2}e^{i\delta_2}, \qquad
\bar A = A_1e^{-i\phi_1}e^{i\delta_1} + A_2e^{-i\phi_2}e^{i\delta_2}.
\eeq
To first order in $r \equiv A_2/A_1$ 
Eq.~(\ref{aCP}) reduces to \cite{Gronau}
\beq
a_{f_{CP}}^{cos} = -[2r\sin(\phi_{12})\sin(\delta_{12})]
\label{acos}
\eeq
and
\beq
a_{f_{CP}}^{sin} = -[\sin 2(\phi_M + \phi_1) + 2r\cos 2(\phi_M + \phi_1)
                   \sin(\phi_{12})\cos(\delta_{12})]
\label{asin}
\eeq
where we have defined 
$\phi_{12}=\phi_1-\phi_2$ and $\delta_{12}=\delta_1-\delta_2$.

In the case $r=0$ or $\phi_{12}=0$
one recovers the case studied above, where
$a_{f_{CP}}^{sin}$ cleanly measures the $CP$ violating quantity $\sin
2(\phi_M + \phi_1)$. If $r \ne 0$ and $\phi_{12}\ne 0$ 
we can consider 2 distinct scenarios:

$(a)$ Direct $CP$ violation ($a_{f_{CP}}^{cos} \ne 0$). 
This occurs when $\delta_{12} \ne 0$ and can be measured by a careful
study of the time dependence since it gives rise to a $\cos\Delta Mt$
term in addition to the $\sin \Delta Mt$ term. Such a scenario would
also give rise to $CP$ asymmetries in charged $B$ decays.

$(b)$ Different quark level 
decay channels that measure the same phase when only one
amplitude contributes, can measure different phases if more than one
amplitude contributes, \ie\ two different processes with the same
$\phi_1$, but with different $r$ or $\phi_2$.

For the rest of this lecture we concentrate on the information we can
get from $a_{f_{CP}} ^{sin}$. To this end we write
\beq
a_{f_{CP}}^{sin} \equiv a_{f_{CP}} = -\sin 2(\phi_0 + \dphi),
\label{aa}
\eeq
where $\phi_0 = \phi_M + \phi_1$, and
$\dphi$ is the correction to it. For small $r$, $\dphi \le r$. However
for $r > 1$, $\dphi$ can take any value. Thus, when we 
catalog values of $\delta\phi$ for various models, 
we use $\delta\phi\simeq 1$ to indicate an arbitrary value.


\section{The Standard Model}

All the information about flavor and $CP$ violation in the \sm\ is
encoded in the CKM matrix. Although the CKM matrix could have up 
to five large phases (only one of which is independent), 
we know experimentally that only two of these are large.
Thus we can write the CKM matrix as:
\beq
V_{CKM}=\left(\begin{array}{lll}
        V_{ud} & V_{us} & |V_{ub}|e^{-i\gamma} \\
        V_{cd} & V_{cs} & V_{cb} \\
	|V_{td}|e^{-i\beta} & V_{ts} & V_{td}
\end{array}\right)
\eeq
The phase strucutre and the magnitudes of the elements are most
transparent in the Wolfenstein parametrization \cite{Wolfenstein}, 
where the CKM matrix is given by
\beq
V_{CKM}=\left(\begin{array}{ccc}
        1-\frac{1}{2}\lambda^2 & \lambda & A\lambda^3(\rho-i\eta)\\
         -\lambda & 1-\frac{1}{2}\lambda^2 & A\lambda^2 \\
         A\lambda^3(1-\rho-i\eta) & -A\lambda^2 & 1 \end{array}\right)
\eeq
here $\lambda = 0.22$ is the Cabbibo angle.
Unitarity of the CKM matrix implies the relation 
\beq
V_{cd}V_{cb}^*+V_{ud}V_{ub}^*+V_{td}V_{tb}^*=0
\eeq
which is usually graphically represented as the 'unitarity triangle'
in the $\rho-\eta$ plane (see Fig. 1).
In principle, one can determine $\beta$ and $\gamma$ 
(or alternatively $\rho$ and $\eta$) from the 
available data on $K$ and 
$B$ decays. However, given the large theoretical uncertainties in the input
parameters (\eg\ $B_K$, $f_B$) the size of these 
phases remains uncertain (for recent reviews see
Refs. \citenum{gnalb,Bur-Fl}). 

\epsfxsize=306pt
\epsfbox{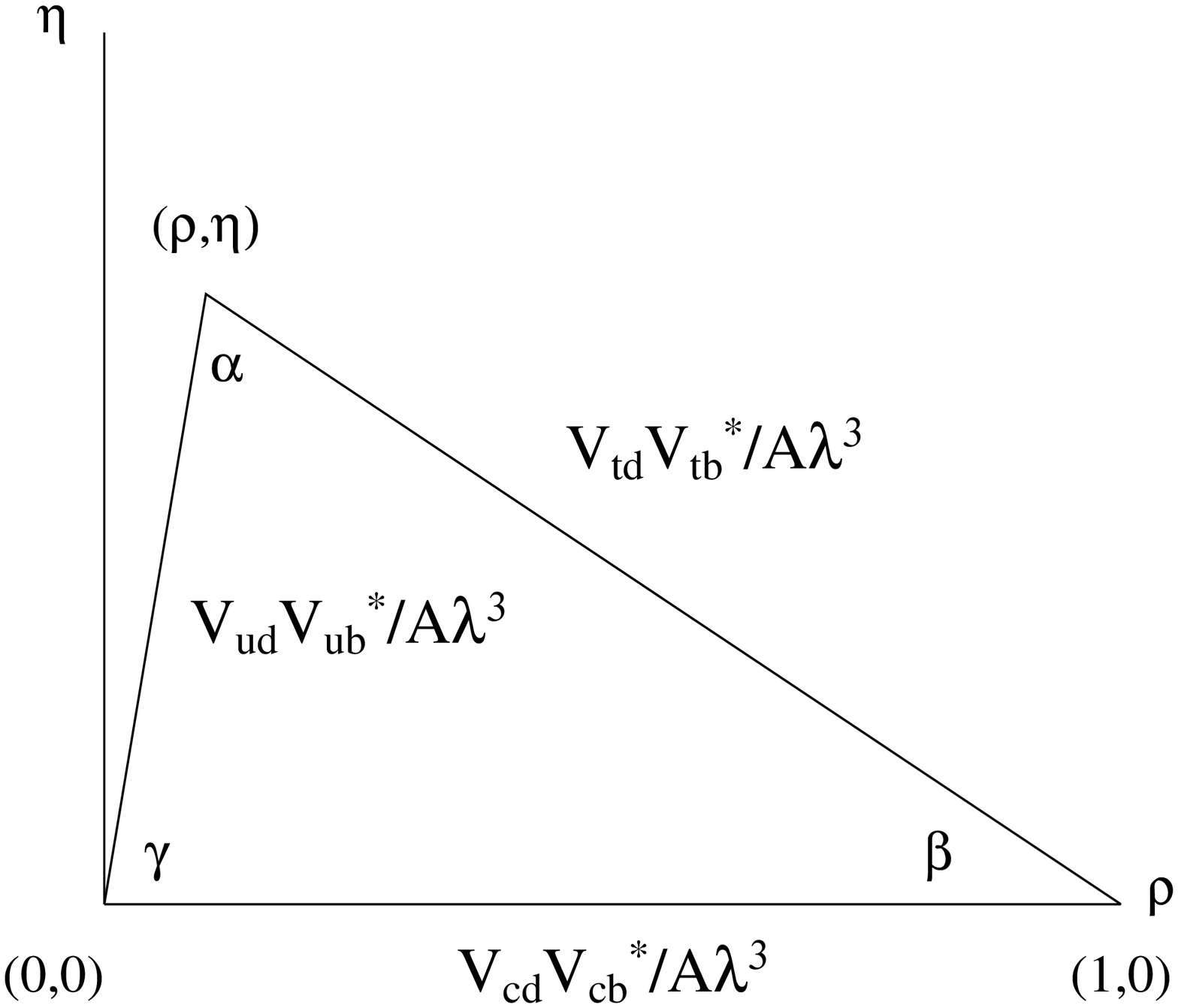}
{\rm Figure 1. }
{The Unitarity Triangle.}
\bigskip
\bigskip

This is where the $CP$ violating experiments at the $B$ factories come
into their own. In the \sm, the \bbbar\ mixing amplitude is dominated
by the box diagram with top quarks in the loop. Thus,
the phase of the mixing amplitude
is given by the phase of $(V_{tb}V_{td}^*)^2$ and in the
convention for the CKM matrix above, we get $\phi_M = 2\beta$. 
In order to extract the CKM phases, 
we then need decay modes of the $B$'s that are dominated by
one decay amplitude, depend on independent CKM phases and are
experimentally feasible. Some examples are:
\newpage
\begin{description}
\item[$(i)$]
{$B \to \psi K_S$:\cite{Ca-Sa} 
The decay is driven by the quark level process
$b\to c\bar c s$. Moreover, the 
dominant contribution to $K-\bar K$ mixing is proportional to
$V_{cs}V_{cd}^*$  (the box diagram with charm quarks).
Thus the CKM elements in the decay amplitude are
$(V_{cb}^*V_{cs})(V_{cs}^*V_{cd})$ leading to $\phi_1=0$ and subsequently 
$\apks = \sin 2\beta$. This decay has a high rate, 
$BR[B\to\psi K_S]=4\times 10^{-4}$ with the $\psi$ tagged by its decay
into 2 leptons, $BR[\psi\to l^+l^-]=0.12$. Moreover there is
negligible pollution from sub-leading decay amplitudes.}
\item[$(ii)$]
{$B \to \pi^+\pi^-$: This decay gets a tree-level contribution from
the quark process $b\to u\bar u d$. 
Thus the CKM elements in the decay amplitude are
$V_{ub}^*V_{ud}$ leading to $\phi_1=\gamma$ and subsequently 
$a_{\pi\pi} = \sin 2(\beta+\gamma)=\sin 2\alpha$. 
The expected rate is $BR[B \to \pi^+\pi^-]\sim 1\times 10^{-5}$.
There is expected to be a
substantial pollution to this prediction coming from the penguin
induced $b\to d \bar u u$ decay. However, it may be possible to still
obtain a measure of $\alpha$ by measuring other isospin related $B\to
\pi\pi$ rates.\cite{Gro-Lon}}
\item[$(iii)$]
{$B \to \phi K_S$:\cite{Lon-Pec} This decay is driven by the quark
level process $b\to s\bar s s$. The leading contribution to this
decay is a penguin diagram with top quarks in the loop. 
Thus the CKM elements in the decay amplitude (after including $K-\bar
K$ mixing) are
$(V_{tb}^*V_{ts})(V_{cs}^*V_{cd})$ leading to $\phi_1=0$ and subsequently 
$\afks = \sin 2\beta$. 
The expected rate is $BR[B\to \phi K_S] \sim 1\times 10^{-5}$, with
the $\phi$ tagged by its decay into two $K$'s: 
$BR[\phi\to K^+K^-]=0.5$. 
As we will discuss later, the $CP$ asymmetry in this mode is 
particularly sensitive to new physics contributions \cite{GrWo},
moreover the \sm\ pollution to
this mode is small and quantifiable \cite{GIW}. Thus this mode
provides an interesting consistency check.}
\end{description}


\section{New Physics in the $B-\bar B$ Mixing Amplitude}

In this section we study the possibility of detecting new
contributions to the \bbbar\ mixing amplitudes.\cite{GNW} We discuss a
construction that allows us to extract information about the CKM
matrix elements, as well as the phase and the magnitude of the new
physics contribution. We highlight potential difficulties in carrying
out this construction, and suggest ways to overcome them.

\newpage

\subsection{The Basic Assumptions and Results}

The first two $CP$ asymmetries to be measured in a $B$ factory
are likely to be
\beq
{\Gamma(B^0_{\rm phys}(t)\to\psi K_S)-
\Gamma(\bar B^0_{\rm phys}(t)\to\psi K_S)\over
\Gamma(B^0_{\rm phys}(t)\to\psi K_S)+
\Gamma(\bar B^0_{\rm phys}(t)\to\psi K_S)}= \apks\sin(\Delta m_B t),
\label{firsta$CP$b}
\eeq
\beq
{\Gamma(B^0_{\rm phys}(t)\to\pi\pi)-
\Gamma(\bar B^0_{\rm phys}(t)\to\pi\pi)\over
\Gamma(B^0_{\rm phys}(t)\to\pi\pi)+
\Gamma(\bar B^0_{\rm phys}(t)\to\pi\pi)}= \app\sin(\Delta m_B t).
\label{firsta$CP$a}
\eeq
%
In addition, the $B$ factory will improve our knowledge of the $B-\bar B$
mixing parameter, $x_d\equiv{\Delta m_B\over\Gamma_{B}}$, and of the
charmless semileptonic branching ratio of the $B$ mesons.

Within the \sm, these four measurements are useful in
constraining the \UT. The asymmetries Eq. (\ref{firsta$CP$b}) and 
Eq. (\ref{firsta$CP$a})
measure angles of the \UT:
\beq\label{SMaPK}
\apks=\ \sin2\beta,
\eeq
\beq\label{SMaPP}
\app=\ \sin2\alpha,
\eeq
where
\beq\label{SMangles}
\alpha\equiv\
\arg\left[-{V_{td}V_{tb}^*\over V_{ud}V_{ub}^*}\right],\ \ \
\beta\equiv\
\arg\left[-{V_{cd}V_{cb}^*\over V_{td}V_{tb}^*}\right].
\eeq
In Eq. (\ref{SMaPK}) we have taken into account the fact that the final
state is $CP$-odd.
In Eq. (\ref{SMaPP}) we have ignored possible penguin contamination 
which can, in principle, be eliminated by isospin analysis
\cite{Gro-Lon}.
The measurement of $x_d$ determines one side of the \UT\ ($R_t$)
upto the unknown constant $\sqrt{B_B}f_B$:
\beq\label{SMxd}
x_d=C_t R_t^2,
\eeq
where
\beq\label{defRt}
R_t\equiv\left|{V_{tb}^*V_{td}\over V_{cb}^*V_{cd}}\right|,
\eeq
and $C_t=\tau_b{G_F^2\over6\pi^2}\eta_B m_B(B_Bf_B^2)m_t^2
f_2(m_t^2/m_W^2)|V_{cb}^*V_{cd}|^2$ (for definitions and notations see
Ref. \citenum{Yossi}).
The present values are $x_d= 0.73 \pm 0.05$ and
$C_t \sim 0.4 - 0.8$ for $\sqrt{B_B}f_B = 140-200$ MeV
(Ref. \citenum{pdg}).
Measurements of various inclusive and exclusive $b\to u\ell\nu$ processes
will determine 
(up to uncertainties arising from various hadronic models) 
the length of
the other side of the \UT\ ($R_u$):
\beq\label{SMbu}
{\Gamma(b\to u\ell\nu)\over\Gamma(b\to c\ell\nu)}=
{1\over F_{{\rm ps}}}\left|{V_{cd}\over V_{ud}}\right|^2 R_u^2,
\eeq
where
\beq\label{defRb}
R_u\equiv\left|{V_{ub}^*V_{ud}\over V_{cb}^*V_{cd}}\right|
\eeq
and $F_{\rm ps}\approx0.5$ is a phase space factor.
The present value for $R_u$ ranges from 0.27 to 0.45 depending on the
hadronic model used to relate the measurement at the
endpoint region, or of some exclusive mode, to the total
$b \to u$ inclusive rate \cite{pdg}.

In the presence of \np\ it is quite possible that the \sm\
predictions Eqs. (\ref{SMaPK},\ref{SMaPP},\ref{SMxd}) 
are violated. The most likely
reason is a new, significant contribution to $B-\bar B$ mixing that
carries a $CP$ violating phase different from the \sm\ one.
Other factors that could affect the construction of the \UT\
from these four measurements are unlikely to be significant
\cite{NiSi,DLN}.
\begin{description}
\item{a.} {The $\bar b\to\bar cc\bar s$ and $\bar b\to\bar uu\bar d$
decays for $\apks$ and $\app$ respectively,
as well as the semileptonic $B$ decays for $R_u$, are mediated by
\sm\ tree level diagrams. In most extensions of the Standard Model
there is no decay mechanism that could significantly compete with
these contributions. (For exceptions, which could affect
the $\bar b\to\bar uu\bar d$ decay see Ref. \citenum{GrWo})}
\item{b.} {New physics could contribute significantly to $K-\bar K$
mixing. However, the small value of $\epsK$ forbids large deviations
from the \sm\ phase of the mixing amplitude. }
%
\item{c.} {Unitarity of the three generation CKM matrix is maintained
if there are no quarks beyond the three generations of the Standard
Model. Even in models with an extended quark sector the effect on
$B-\bar B$ mixing is always larger than the violation
of CKM unitarity.} 
\end{description}

Our analysis below applies to models where the above three conditions are
not significantly violated. Under these circumstances the relevant \np\
effects can be described by two new parameters, $r_d$ and $\theta_d$
\cite{SoWo,DDO,SiWo,CKLN},
defined by
\beq\label{thetad}
\left(r_d e^{i\theta_d}\right)^2\equiv\frac{
\langle{B^0|{\cal H}^{\rm full}_{\rm eff}|\bar B^0}\rangle}
{\langle{B^0|{\cal H}^{\rm SM}_{\rm eff}|\bar B^0}\rangle}
,
\eeq
where ${\cal H}^{\rm full}_{\rm eff}$ is the effective Hamiltonian
including both \sm\ and \np\ contributions, and
${\cal H}^{\rm SM}_{\rm eff}$ only includes the \sm\ box diagrams.
In particular, with this definition, the modification of the two $CP$
asymmetries in Eqs. (\ref{SMaPK},\ref{SMaPP}) 
depends on a {\it single} new parameter, the phase $\theta_d$:
\beq\label{NPaPK}
\apks=\sin(2\beta + 2\theta_d),
\eeq
\beq\label{NPapp}
\app=\sin(2\alpha-2\theta_d),
\eeq
while the modification of the $B-\bar B$ mixing parameter $x_d$
in Eq. (\ref{SMxd}) is given by the magnitude rescaling parameter, $r_d$:
\beq\label{NPxd}
x_d=C_t R_t^2 r_d^2.
\eeq
Furthermore, since the determination of $R_u$ from the
semileptonic $B$ decays is not affected by the \np, and
since the \UT\ remains valid, we have the following
relations between the length of its sides and its angles:
\beq\label{Rbab}
R_u={\sin\beta\over\sin\alpha},
\eeq
\beq\label{Rtab}
R_t={\sin\gamma\over\sin\alpha},
\eeq
where
\beq\label{defgamon}
\gamma\equiv\arg\left[-{V_{ud}V_{ub}^*\over V_{cd}V_{cb}^*}\right].
\eeq
When $\alpha$, $\beta$ and $\gamma$
are defined to lie in the $\{0,2\pi\}$ range, they satisfy
\beq\label{defgam}
\alpha+\beta+\gamma=\pi\ {\rm or}\ 5\pi.
\eeq

The four measured quantities \apks, \app, $x_d$ and $R_u$
can be used to achieve the following:\cite{SoWo}
\begin{description}
\item{(i)}{Fully reconstruct the unitarity triangle and,
in particular, find $\alpha$, $\beta$ and $R_t$;}
\item{(ii)} {Find the magnitude and phase of the \np\ contribution
to $B-\bar B$ mixing, namely determine $r_d$ and $\theta_d$.}
\end{description}

It is straightforward to show that the above tasks are possible.
Eqs. (\ref{NPaPK},\ref{NPapp},\ref{Rbab}) 
give three equations for three unknowns,
$\alpha$, $\beta$ and $\theta_d$. Once $\alpha$ and $\beta$ are known,
$\gamma$ can be extracted from Eq. (\ref{defgam}), 
$R_t$ can then be deduced from
Eq. (\ref{Rtab}), and finally $r_d$ is found from Eq. (\ref{NPxd}).

In the next two sub-sections we describe how to determine the
parameters, both in the $\rho-\eta$ plane, and in the
$\sin 2\alpha-\sin 2\beta$ plane.
In practice, however, it is quite likely that the combination of
experimental and theoretical uncertainties (particularly in
the $x_d$ and $R_u$ constraints) and discrete ambiguities
will limit the usefulness of the above method rather significantly.
We discuss the source of the hadronic
uncertainties in Sec. 3.4 and the discrete ambiguities that arise in
this calculation in Sec. 3.5. We mention ways to
resolve some of the ambiguities in the sub-section 3.6.


\subsection{The $\rho-\eta$ Plane}

The key point in the extraction of the CKM parameters is that the
angle $\theta_d$ cancels in the following sum:
\beq\label{indsum}
2(\alpha+\beta)=\arcsin(\apks)+\arcsin(\app).
\eeq
In other words, the angle $\gamma$ can be determined (up to the
discrete ambiguities to be discussed in subsection 3.5). 
In the $\rho-\eta$ plane, a value
for $\gamma$ gives a ray from the origin, while a value for $R_u$ gives
a circle that is centered in the origin. The intersection point
of the line and the circle gives $(\rho,\eta)$ of the unitarity
triangle and determines it completely.

A graphical way to carry out these calculations in the $\rho-\eta$ plane
is the following (see Figure 2) \cite{CKLN}. One draws the four curves that
correspond to Eqs. (\ref{SMaPK},\ref{SMaPP},\ref{SMxd},\ref{SMbu}) 
(even though only Eq. (\ref{SMbu}) is valid in the presence of new 
physics). 
The next step is to draw the ray from the origin
that passes through the intersection point of the $\beta$-ray and the
$\alpha$-circle: this is the {\it correct} $\gamma$-ray (see the dashed
line in Figure 2). The intersection point of the $\gamma$-ray and the
$R_u$-circle gives the {\it correct} vertex of the \UT, $(\rho,\eta)$,
namely
\beqa\label{trueUT}
\tan\beta=&\ {\displaystyle{\eta\over1-\rho}},\nonumber \\
R_t^2=&\ \eta^2+(1-\rho)^2.
\eeqa
The information about the \np\ contribution to $B-\bar B$ mixing
is found from the intersection point of the $\beta$-ray and the
$x_d$-circle, ($\rho',\eta'$), namely
\beqa\label{findtdRd}
\theta_d=&\ {\displaystyle{
\arctan{\eta'\over1-\rho'}-\arctan{\eta\over1-\rho}}},\nonumber \\
r_d^2=&\ {\displaystyle{{\eta'^2+(1-\rho')^2\over \eta^2+(1-\rho)^2}.}}
\eeqa

\epsfxsize=356pt
\epsfbox{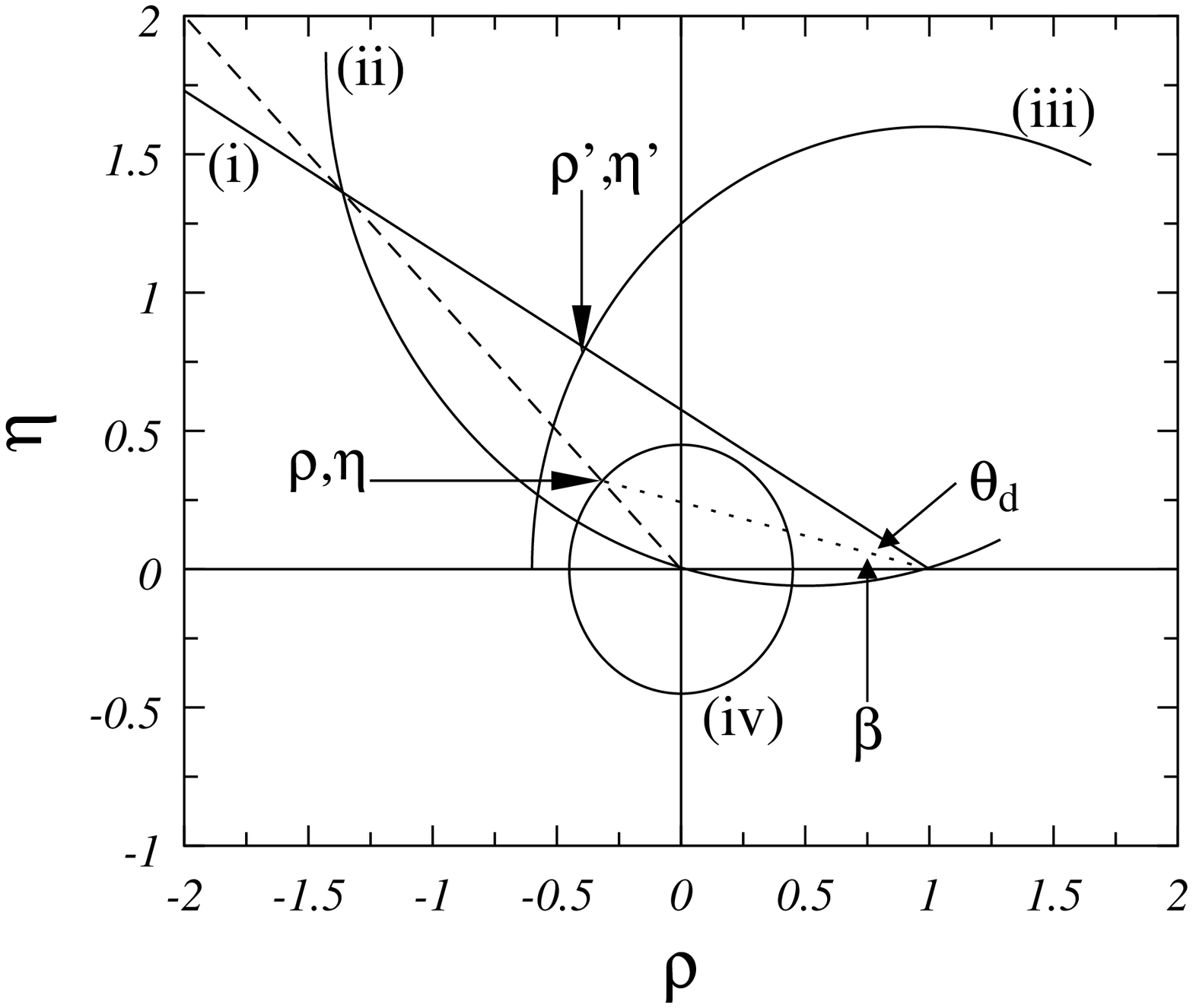}
{\rm Figure 2.}
{The model independent analysis in the $\rho-\eta$ plane:
 (i) The $\apks$ ray; (ii) The $\app$ circle;
 (iii) The $x_d$ circle; (iv) The $R_u$ circle.
 The $\gamma$ ray is given by the dashed line.
 The true $\beta$ ray is given by the dotted line.
 Also shown are the true vertex of the \UT\ ($\rho,\eta$) and
 the ($\rho',\eta'$) point that serves to find $\theta_d$ and $r_d$.}
\newpage

\subsection{The $\sin2\alpha-\sin2\beta$ Plane}

A presentation of the various constraints in the $\sin2\alpha-\sin2\beta$
plane
\cite{SoWo,NiSa,GrNibsg} is useful because the two angles are usually
correlated. \cite{DDGN}
The model independent analysis is demonstrated in Figure 3. The
$R_u$ constraint gives an eight-shaped curve on which the physical
values have to lie.
The various solutions for Eq. (\ref{indsum}) fall on two ellipses, the
intersections of which with the $R_u$ curve determine the allowed
values of $\sin2\alpha$ and $\sin2\beta$.
Note that these ellipses cross the eight-shaped curve in sixteen points
but, as argued above, only eight of these points are true solutions.
The inconsistent intersection points can be found by noting that the
slopes of the ellipse at the consistent points should be
$(\cos2\alpha,-\cos2\beta)$. The eight correct solutions are
denoted by the filled circles in Figure 3.

In the above, we showed how to use measured values of the $CP$ asymmetries
\apks\ and \app\ to find the allowed values for $\alpha$ and $\beta$.
The presentation in the $\sin2\alpha-\sin2\beta$ plane is also useful
for the opposite situation. Some models predict specific values for
$\alpha$ and $\beta$. (Such predictions can arise naturally from
horizontal
symmetries.) On the other hand, the models often allow new contributions
to $B-\bar B$ mixing of unknown magnitude and phase. In this case, the
predicted value of $(\sin2\alpha,\sin2\beta)$ is just a point in the
plane, and the ellipse Eq. (\ref{indsum}) actually gives the allowed (and
correlated) values of $(\app,\apks)$. (Such an analysis was carried out in
Ref. \citenum{BHR}).

More generally, even in models that make no specific predictions
for CKM parameters, we usually have some constraints on the allowed
range for $\alpha$ and $\beta$. For example, in this work we assume
the validity of the limits on $R_u$ from charmless semileptonic
$B$ decays which constrains the ratio $\sin\beta/\sin\alpha$ through
Eq. (\ref{Rbab}). 
Note, however, that this constraint by itself cannot exclude
any region in the $\app-\apks$ plane. The reason is the following.
For any value of $R_u$, neither $\alpha$ nor $\theta_d$ are constrained.
(The angle $\beta$ is constrained for any $R_u<1$ and certainly by
the present range, $0.27<R_u<0.45$.)
Then any value of $\apks$ can be accommodated by an appropriate
choice of $\theta_d$ and any value of $\app$ can be fitted by further
choosing an appropriate $\alpha$. Obviously, to get predictions
for the $CP$ asymmetries beyond the Standard Model, one has to make
some assumptions that go beyond our generic analysis.

For example, consider models where $\epsK$ is dominated
by the \sm\ box diagrams (while $B-\bar B$ mixing is not). Then, we know
that $0<\gamma<\pi$. This already excludes part of the allowed
range. In particular, $(\app,\apks)=(1,-1)$ or $(-1,1)$ requires
$\gamma=0$ or $\pi$, and is therefore excluded in this class of
models. More generally, in any class of models where $\sin^2\gamma$
cannot assume any value between zero and one,
some regions in the \app-\apks\ plane are excluded.

\epsfxsize=306pt
\epsfbox{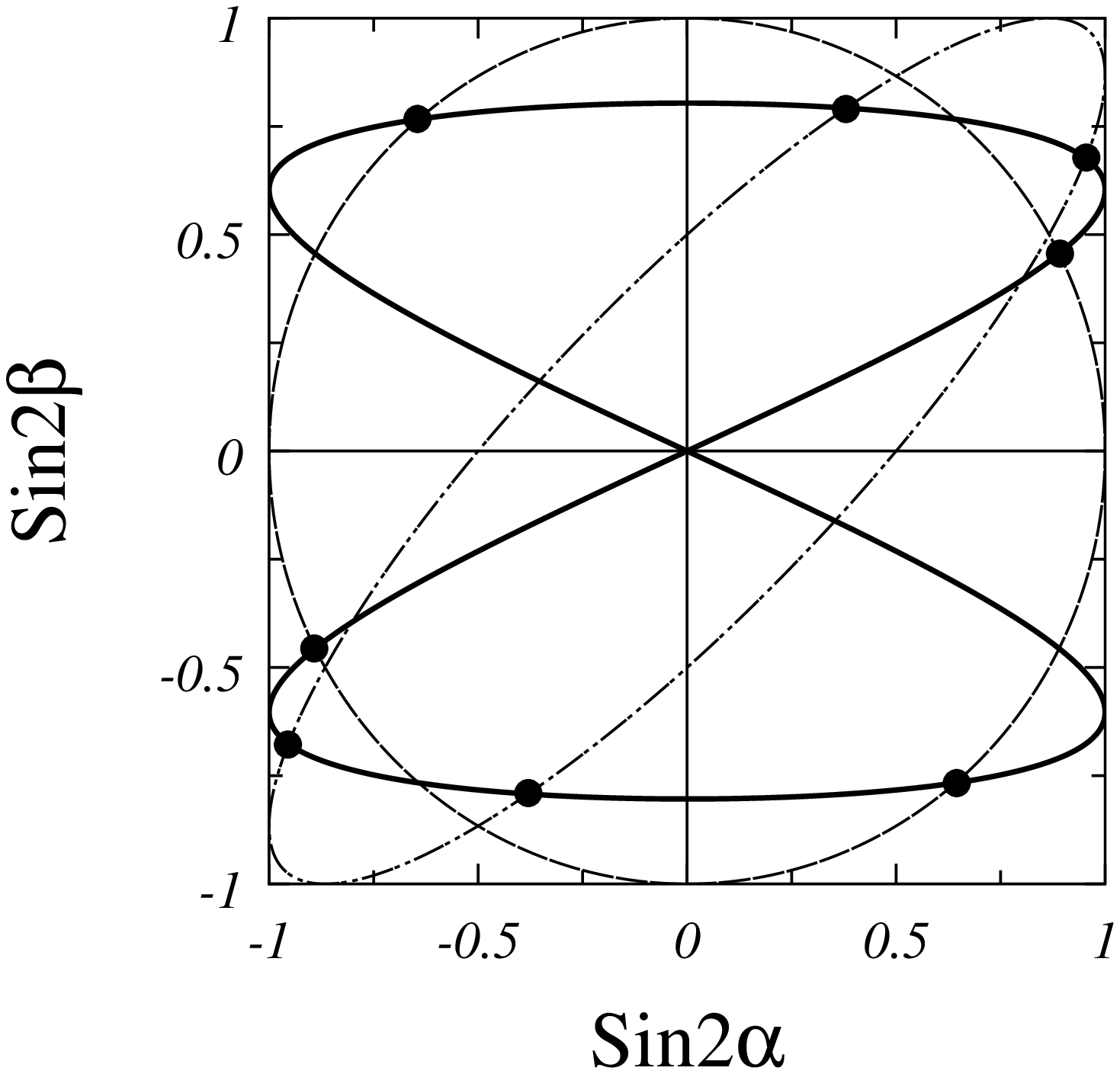}
{\rm Figure 3. }
{The $\alpha+\beta$ constraint Eq. (\ref{indsum}) 
and the $R_u$ constraint Eq. (\ref{Rbab})
in the $\sin 2\alpha-\sin 2\beta$ plane. The eight possible solutions
for the \UT\ are given by the filled circles.}
\newpage


\subsection{Hadronic Uncertainties}

In Sec. 2 we argued that the $CP$ asymmetries in $B$ decays
that are a result of interference between mixing and decay give us a
clean measurement of $CP$ violating quantities that are free of hadronic
uncertainties if only one decay amplitude contributes. Yet in the
presence of new contributions to the \bbbar\ mixing amplitude, we find
we are once again limited by our theoretical understanding of hadronic
physics. To understand the source of the hadronic uncertainty, it is
instructive to compare the $CP$ violation in neutral $B$ 
decays to $CP$ eigenstates with that in neutral $K$ decays to $CP$
eigenstates. 

The $CP$ violation in the decay $K_L \to \pi\pi$ is also a result of
interference between mixing and decay. The quark level decay is given
by the process $s \to u\bar u d$ with CKM matrix elements
$V_{us}^*V_{ud}$ which are real in the convention we have chosen.
Thus, as argued above, the $CP$ asymmetry in this mode cleanly
measures $\sin 2\phi_{M_K}$, the phase of the $K-\bar K$ mixing
amplitude. The problem arises in trying to relate
$\phi_{M_K}$ to phases of CKM matrix elements. 
This is because although the decay was dominated by one
contribution, in this case the mixing amplitude has more than one
contribution with unknown relative magnitudes and different (but
known) dependence on CKM matrix elements. In particular, there is a
large, unknown, 
long-distance contribution to the $K-\bar K$ mixing amplitude, making
the interpretation in terms of CKM parameters dependent on poorly
known hadronic quantities like $B_K$.

Similarly, in the presence of new, unknown, contributions to the
\bbbar\ mixing amplitude, although \apks\ cleanly measures 
$\sin 2\phi_{M_B}$, it is not possible to relate this to fundamental
parameters like the CKM matrix elements without a knowledge of the
relative magnitudes of the different contributions. The clean
information we had before is lost, and the extraction of CKM
parameters is once again dependent on hadronic parameters like
$B_Bf_B^2$ that are not well determined at present. 

\subsection{Discrete Ambiguities}

A major obstacle in carrying out the above program will be the discrete
ambiguities in determining $\gamma$. We now describe these ambiguities.

A physically meaningful range for an angle is $2\pi$.
We choose this range to be $\{0,2\pi\}$.
Measurement of any single asymmetry,
$\sin2\phi$, determines the corresponding angle only up to a fourfold
ambiguity: $\phi$, $\pi/2-\phi$, $\pi+\phi$ and $3\pi/2-\phi$ (mod $2\pi$).
Specifically, let us denote by $\bar\alpha$ and $\bar\beta$
some solution of the equations
\beq\label{somesol}
\apks=\sin2\bar\beta,~~~~\app=\sin2\bar\alpha.
\eeq
Thus, measurements of the two asymmetries leads to a sixteenfold
ambiguity in the values of the $\{\bar\alpha,\bar\beta\}$ pair.
However, since $\bar\alpha=\alpha-\theta_d$ and $\bar\beta=\beta+\theta_d$, and
unitarity is not violated, $\gamma$ still satisfies the condition
\beq\label{NPsum}
\bar\alpha +\bar\beta +\gamma=\pi~ (mod~ 2\pi).
\eeq
Then, the sixteen possibilities for $\gamma$ are divided into
two groups of eight that are related by the combined operation
$\bar\alpha \to \bar\alpha +\pi$ and $\bar\beta \to \bar\beta +\pi$.
This, in turn shifts the value of $\gamma$ by $2\pi$.
However, since $\gamma$ is only defined modulo $2\pi$,
the ambiguity in $\gamma$ is reduced to eightfold.
We emphasize that this reduction of the ambiguity
depends only on the definition of $\gamma$. Defining
\beq\label{defphipm}
\phi_\pm=\bar\alpha\pm\bar\beta,
\eeq
the eight possible solutions for $\gamma$ are
\beq\label{MIgamma}
\gamma=\pm\phi_+,\ \pi\pm\phi_+,\ \pi/2\pm\phi_-,\
3\pi/2\pm\phi_-\ (mod~ 2\pi).
\eeq
Note that the eight solutions come in pairs of $\pm\gamma$.
This in turn implies that the ambiguity on $R_t$ is only fourfold.

In any model where the three angles $\bar\alpha$, $\bar\beta$, \
and $\gamma$ form a
triangle, the ambiguity is further reduced
\cite{Yossi}:
the requirement that the angles are either all in the range
$\{0,\pi\}$ or all in the range $\{\pi,2\pi\}$ reduces the ambiguity
in $\gamma$ to fourfold. It is enough to know the signs of $\apks$ and
$\app$ to carry out this step. Finally, within the \sm, the bound
$0<\beta<\pi/4$ (obtained from the sign of $\epsK$ and from
$R_u < 1/\sqrt{2}$) reduces the ambiguity in $\gamma$ to twofold.

When we allow for the possibility of \np\ effects in the mixing,
knowing the signs of \apks\ and \app\ does not lead to further reduction
in the ambiguity, which remains eightfold.
The three angles $\bar\alpha$, $\bar\beta$ and
$\gamma$ are not angles that define a triangle and therefore further
constraints cannot be imposed. It is possible, for example, that
both $\gamma$ and $\bar\beta$ lie in the range $\{\pi/2,\pi\}$.
Further the sign of $\epsK$ may not be related to the sign of $\eta$.

The following example will make the situation clear. Take
\beq\label{examplea}
\app=1/2,\ \ \ \apks=\sqrt3/2.
\eeq
Then, we could have
\beq\label{apria}
\bar\alpha={\pi\over12},{5\pi\over12},{13\pi\over12},{17\pi\over12},\ \
\bar\beta={\pi\over6},{\pi\over3},{7\pi\over6},{4\pi\over3}.
\eeq
The eight solutions for $\gamma$ are
\beq\label{solveaMI}
\gamma={\pi \over 4},\ {5\pi\over 12},\ {7\pi\over 12},\
{3\pi\over 4},\
{5\pi\over 4},\ {17\pi\over 12},\ {19\pi\over 12},\ {7\pi\over 4}.
\eeq
%
If $\bar\alpha,\bar\beta,\gamma$
define a triangle, then only four solutions are allowed:
\beq\label{solvea}
(\bar\alpha,\bar\beta,\gamma)=
\left({\pi\over12},{\pi\over6},{3\pi\over4}\right),
\left({\pi\over12},{\pi\over3},{7\pi\over12}\right),
\left({5\pi\over12},{\pi\over6},{5\pi\over12}\right),
\left({5\pi\over12},{\pi\over3},{\pi\over4}\right).
\eeq
Assuming $0<\bar\beta<\pi/4$ as in the \sm\ leaves only the first
two choices.

In various specific cases, the discrete ambiguity is smaller. If the two
asymmetries are equal in magnitude,
there is only a sixfold ambiguity:
\beqa\label{MIequal}
\app=\apks\ \Longrightarrow&\ \gamma=
\pm2\bar\beta,\ \pi\pm2\bar\beta,\ \pi/2,\ 3\pi/2\ (mod~ 2\pi),\\ \nonumber
\app=-\apks\ \Longrightarrow&\ \gamma=0,\ \pi,\ \pi/2\pm2\bar\beta,\
3\pi/2\pm2\bar\beta\ (mod~ 2\pi).
\eeqa
If one of the asymmetries is maximal, there is a fourfold ambiguity, e.g.
\beqa\label{MIma}
\app=+1\ \Longrightarrow&\ \gamma=
\pm(\pi/4+\bar\beta),\ \ \pm(3\pi/4-\bar\beta)\ \ (mod~
2\pi),\nonumber \\
\app=-1\ \Longrightarrow&\ \gamma=\pm(\pi/4-\bar\beta),\ \
\pm(3\pi/4+\bar\beta)\ \ (mod~ 2\pi).
\eeqa
If both asymmetries are maximal, the ambiguity is twofold.
If the two asymmetries vanish, there is only a fourfold ambiguity:
\beq\label{MI$CP$}
\app=\apks=0\ \Longrightarrow\
\gamma=0,\ \pi/2,\ \pi,\ 3\pi/2.
\eeq
This is an interesting case, because it is predicted by models
with approximate $CP$ symmetry (e.g. in some supersymmetric models
\cite{GNR}.
Only two of the solutions ($0,\pi$) correspond to the
$CP$ symmetric case while in the other two ($\pi/2,3\pi/2$), the zero
asymmetries are accidental.

So far we have ignored the penguin contamination in \app.
The isospin analysis eliminates the penguin contamination only up to
a four fold ambiguity \cite{Gro-Lon}.
Therefore, if the isospin analysis is needed, the ambiguities are
increased.

In addition, for each value of $\gamma$ there are two
possibilities for $\theta_d$ related by $\theta_d \to \theta-d +\pi$.
As long as the \np\ is such that the $\Delta b=2$ operator
that contributes to $B-\bar B$ mixing can be separated into two
$\Delta b=1$ operators the $\theta_d \to \theta_d+\pi$ ambiguity is physical.
Otherwise, it is not physical.


\subsection{Final Comments}

We argued that the most likely effect of \np\ on $CP$ asymmetries
in neutral $B$ decays into $CP$ eigenstates will be
a significant contribution to the mixing. This is because we have
concentrated on decays that are allowed at tree level in the \sm. Thus
the \np\ effects on
the decay amplitudes and on CKM unitarity can be neglected in a large
class of models.%
\footnote{%
The \np\ effects may significantly alter the patterns of $CP$ asymmetries
in decays that are dominated by penguins in the \sm\ \cite{GrWo}.
See Sec. 4 for a discussion of this point.}
We explained that in this class of models, the
unitarity triangle can be constructed model independently and the
\np\ contribution to the mixing can be disentangled from the
Standard Model one.

However, the combination of hadronic uncertainties and discrete
ambiguities puts serious obstacles in carrying out this calculation.
In particular, there is an eightfold ambiguity in the construction of
the triangle. In order to
get useful results, it will be necessary to reduce 
the hadronic uncertainties and discrete ambiguities.

One way to eliminate some of the allowed solutions
can be provided by a rough knowledge of
$\cos(2\alpha-2\theta_d)$, $\cos(2\beta+2\theta_d)$ or $\cos2\gamma$
\cite{GrQu}.
For example, $\cos(2\alpha-2\theta_d)$ can be determined
from the $CP$ asymmetry in $B\to\rho\pi$
\cite{SnQu}
$\cos2\gamma$ from $B\to DK$
\cite{GrWy}
While a precise measurement of either of these is not expected
in the first stages of a $B$ factory, a knowledge of the sign
of the cosine is already useful for our purposes: knowing either of
sign[$\cos2(\alpha-\theta_d)$], sign[$\cos2(\beta+\theta_d)$] or
sign[$\cos2\gamma$] reduces the
ambiguity in $\gamma$ to fourfold. Knowing two of them reduces it to
twofold. (Knowing the three of them, however,
cannot be combined to completely eliminate the ambiguity.)

The ambiguity associated with the isospin analysis can be removed
by measuring the time dependent $CP$ asymmetry in $B \to \pi^0\pi^0$ 
\cite{Gro-Lon}.
Another way is by studying
$B\to\rho\pi$ \cite{SnQu,GrQu}. Here, due to interference between
several amplitudes, isospin relations can be used to determine
$\sin 2\alpha$ without penguin contamination, and without any discrete
ambiguity. 

A different approach is to make further assumptions about the
\np\ that is responsible for the effects discussed above.
For example, in the \sm, there is a strong correlation between \apks\
and $\apnn\equiv 
\Gamma(K_L\to\pi^0\nu\bar\nu)/\Gamma(K^+\to\pi^+\nu\bar\nu)$
\cite{Bur-Buc}, which we illustrate in Fig. 4. However,
in most supersymmetric models processes involving third
generation quarks, such as $B-\bar B$ mixing, are significantly
modified by the \np, but processes with only light quarks,
such as $K\to\pi\nu\bar\nu$, are not.\cite{Nir-Worah} 
Thus finding $(\apks,\apnn)$ outside the allowed region in Fig. 4
would most likely be due to new physics in the \bbbar\ mixing
amplitude.
Then measurements of
$K^+\to\pi^+\nu\bar\nu$ and $K_L\to\pi^0\nu\bar\nu$ will provide
the true values of $R_t$ or $|\eta|$, respectively.
Although one could construct contrived supersymmetric models with large
contributions to the $K\to\pi\nu\bar\nu$ decays, this possibility is 
often signalled by large, observable $D-\bar D$ mixing.\cite{Nir-Worah}
The unitarity triangle can be determined from these
up to a fourfold ambiguity. The additional input of $R_u$
reduces this to a twofold ambiguity.
The determination of
$\gamma$ by the methods described above will provide a {\it test} of this
class of models. It will not resolve the twofold ambiguity.

\epsfxsize=306pt
\epsfbox{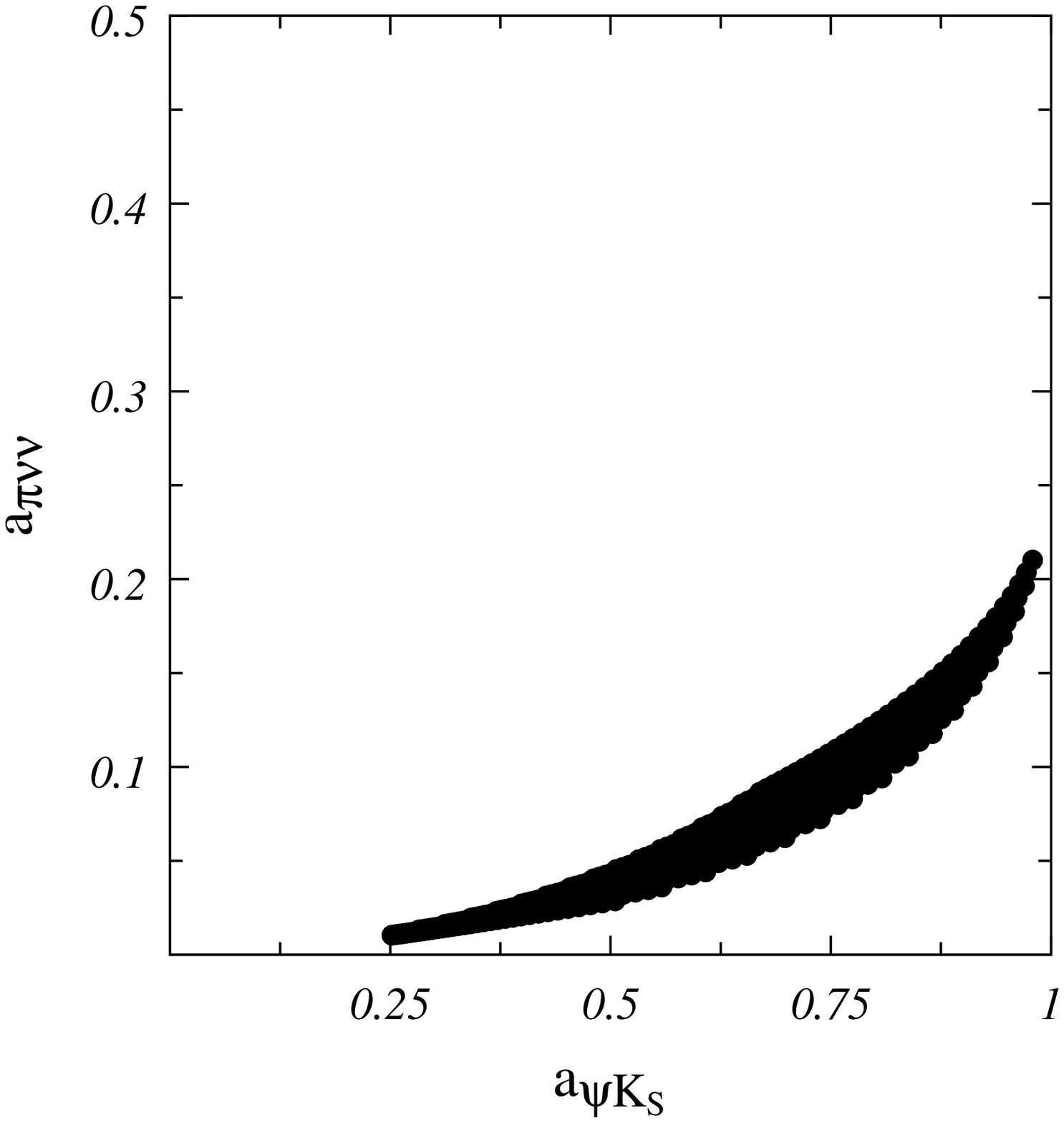}
{\rm Figure 4. }
{The Standard Model allowed region in the \apks - \apnn\ plane. 
We have used $-0.25\le\rho\le 0.40$, $0.16\le\eta\le 0.50$.}

In some models \cite{Nirxs}
there is a significant contribution to both $B_d$ and $B_s$ mixing but
the ratio between the two obeys the Standard Model relation,
\beq\label{BsBd}
{\Delta m_{B_d}\over\Delta m_{B_s}}=F_{SU(3)}\sin^2\theta_C
R_t^2,
\eeq
where $F_{SU(3)}$ is an $SU(3)$-isospin breaking parameter.
Then, a measurement of $\Delta m_{B_s}$ will provide the correct $R_t$
and, again, the unitarity triangle can be determined, up to a twofold
discrete ambiguity, from $R_u$ and $R_t$. The determination of
$\gamma$ by our analysis is in this case, again, a test and
will not resolve the twofold ambiguity. Note, however, that in most
models where the {\it ratio} between $B_d$ and $B_s$ mixing obeys
Eq. (\ref{BsBd}), the {\it phases} in the $B_s,B_d$ mixing amplitudes are
the same as in the \sm, namely $\theta_d=0$. Then $r_d$ is the only new
parameter, and the whole analysis becomes trivial.

In a large class of models, $\epsK$ has only small
contributions from \np. If dominated by the Standard Model,
$\epsK$ implies that all angles of the unitarity triangle are in the
range $\{0,\pi\}$, and the ambiguity is reduced to fourfold.

Of course, one can combine several of these measurements and
assumptions to get a better handle on the true form of the
unitarity triangle. It is obvious however that the model independent
construction of the triangle, while possible in principle, will
pose a serious theoretical and experimental challenge.


\section{New Physics in the $B$ Decay Amplitudes}

\subsection{Introduction}

In this section we 
make a systematic analysis of the effects of new physics in
the $B$ decay amplitudes on the $CP$ asymmetries in 
neutral $B$ decays.\cite{GrWo}
Although these are expected to be smaller than new physics effects 
on the mixing amplitude, they are easier to probe in some cases.
This is based on the fact, that given the current uncertainties in the
values of the CKM phases, the only precise predictions concerning the
$CP$ asymmetries made by the \sm\ are the following:
\begin{description}
\item[$(i)$]
{The $CP$ asymmetries in all $B_d$ 
decays that do not involve direct $b \to u$ (or $b \to d$)
transitions have to be the same.}
\end{description}
This prediction holds for the $B_s$ system in an even stronger form
\begin{description}
\item[$(ii)$]
{The $CP$ asymmetries in all $B_s$ 
decays that do not involve direct $b \to u$ (or $b \to d$) 
transition not only have to be the same, but also approximately vanish.}
\end{description}
Thus, the best place to look for evidence of new $CP$ violating
physics is obviously the $B_s$ system \cite{Nir-Sil,yuvalBs}. 
The $B$ factories, however,
will initially take data at the $\Upsilon(4s)$ where 
only the $B_d$ can be studied.

New physics could in principle contribute to both the mixing matrix and
to the decay amplitudes. As discussed in the previous section, 
it is plausible that the new contributions to the mixing could
be of the same size as the \sm\ contribution since it 
is already a one-loop effect.
This is why most of the existing studies on the
effects of new physics on $CP$ violating $B$ meson decays have
concentrated on effects in the mixing matrix, and assume the decay
amplitudes are those in the \sm\ \cite{Yossi,other,dutta,worah} 
(in Ref. \citenum{dutta} 
a more general analysis was done where they allow for 
new contributions to the penguin dominated \sm\ decay amplitudes).
The distinguishing feature of new
physics in mixing matrices is that its effect is universal, \ie\ 
although it changes the magnitude of the asymmetries it
does not change the patterns predicted by the \sm. 
Thus, the best way to search for these effects
would be to compare the observed $CP$ asymmetry in a particular
decay mode with the asymmetry predicted in the \sm. 
This is straightforward for the leading $B_s$
decay modes where the \sm\ predicts vanishing $CP$ asymmetries.
However, due to the large uncertainties in the \sm\ predictions
for the $B_d$ decays, these new effects would have to be
large in order for us to distinguish them from the \sm. As discussed
in the previous section one would require additional measurements in
order to reduce the hadronic uncertainties and discrete ambiguities
that make these effects difficult to detect. 
In any case, the \sm\ prediction $(i)$ concerning $B_d$ decays
still holds.

In contrast, the effects of new physics in decay amplitudes are
manifestly non-universal, \ie\ they depend on the specific process and
decay channel under consideration. Experiments 
on different decay modes that would measure
the same $CP$ violating quantity in the absence of new contributions
to decay amplitudes, now actually measure different $CP$ violating
quantities. Thus, the \sm\ prediction $(i)$,
concerning $B_d$ decays, can be violated.
Even though the possibility of new physics in decay
amplitudes is more constrained than that in mixing amplitudes,
one could detect these smaller
effects by exploiting the fact that now one does not care about the
predicted value for some quantity, only that two experiments that
should measure the same quantity, in fact, do not.
It is this possibility that we wish to study in this section.

We first discuss the possible decay
channels, and the uncertainties in the universality predictions
introduced within the \sm\ itself by sub leading effects.
To this end we pay special attention to the decay $B\to \phi K_S$
mediated by the neutral current process $b \to s\bar s s$. 
We explain it's
usefulness in probing for new physics, and discuss the possibility of
unexpected long-distance effects
polluting this sensitivity. We propose an experimental test to
constrain this \sm\ pollution. Finally we present a brief
study of models of new physics that could contain new $CP$ violating
decay amplitudes, and their expected size. 


\subsection{The Different Decay Channels}

There are 12 different 
hadronic decay channels for the $b$ quark: 8 of them are
charged current mediated
\beqa
(c1)~ b \to c \bar c s\,, \qquad (c2)~ b \to c \bar c d\,, \qquad &&
(c3)~ b \to c \bar u d\,, \qquad (c4)~ b \to c \bar u s\,, \nonumber \\
(c5)~ b \to u \bar c d\,, \qquad  (c6)~ b \to u \bar c s\,, \qquad &&
(c7)~ b \to u \bar u d\,, \qquad  (c8)~ b \to u \bar u s\,, 
\eeqa
and 4 are neutral current
\beq
(n1)~b \to s \bar s s\,, \qquad  (n2)~ b \to s \bar s d\,, \qquad 
(n3)~b \to s \bar d d\,, \qquad  (n4)~ b \to d \bar d d\,.
\eeq
If only one \sm\ decay amplitude dominates all
of these decay channels, \ie\ $r = 0$ in Eq.~(\ref{asin}),
then up to 
${\cal O}(\lambda^2)$ (where $\lambda \approx 0.22$ is the
expansion parameter in the Wolfenstein approximation), 
the $CP$ asymmetries in $B$ meson decays all measure one of the 4
phases, 
\beqa  \label{SMfour} &&
\alpha \equiv \arg\left(-{V_{td}V_{tb}^*\over
V_{ud}V_{ub}^*}\right), \qquad
\beta \equiv \arg\left(-{V_{cd}V_{cb}^*\over
V_{td}V_{tb}^*}\right),\nonumber \\ &&
\gamma \equiv \arg\left(-{V_{ud}V_{ub}^*\over
V_{cd}V_{cb}^*}\right), \qquad
\beta' \equiv \arg\left(-{V_{cs}V_{cb}^*\over
V_{ts}V_{tb}^*}\right) \simeq 0.
\eeqa 
This situation is nicely summarized, along with relevant decay modes
in Table 1 of \cite{helen-pdg}.
Note that $\beta'< 2.5 \times 10^{-2}$ is very small in the 
SM \cite{Bur-Fl}, 
but in principle measurable. 
For our purpose, however, this small value is a sub-leading correction to
the clean SM prediction $(ii)$. 
We will study corrections to this idealized limit, as well as to the
$r=0$ limit, in the next sub-section.
We now discuss the
effects that new physics in $b$ quark decay amplitudes could have on
the predictions of Eq.~(\ref{SMfour}).

In the \sm\ the $CP$ asymmetries in the decay 
modes $(c1)~b \to  c \bar c s$ (\eg\ $B_d \to
\psi K_S$, $B_s \to D_s^+ D_s^-$), 
$(c2)~b \to  c \bar c d$ 
(\eg\ $B_d \to D^{+}D^{-}$, $B_s \to \psi K_S$), and 
$(c3)~ b \to c \bar u d$ 
(\eg\ $B_d \to D^0_{CP}\rho$, $B_s \to D^0_{CP} K_S$)
all measure the angle $\beta$ in $B_d$ decay
and $\beta'$ in $B_s$ decays.
[$(c5)~b \to u \bar c d$ acts as a correction to $(c3)$ and will be
addressed later].
In the presence of new contributions to the $B-\bar B$ mixing matrix,
the $CP$ asymmetries in these modes would no longer be measuring the
CKM angles $\beta$ and $\beta'$. However, they would all still measure
the same angles $(\beta + \delta_{m_d}, \beta' + \delta_{m_s})$, 
where $(\delta_{m_d},\delta_{m_s})$ are the new contributions to the 
$B_{(d,s)}-\bar B_{(d,s)}$ mixing phase.
In contrast, new contributions to the $b$ quark decay amplitudes 
could affect
each of these modes differently, and thus they would each
be measuring different $CP$ violating quantities. 

Several methods \cite{GrWy} have been proposed
based on the fact that the two amplitudes 
$(c4)~b \to c \bar u s$ 
and $(c6)~b \to u \bar c s$ (\eg\ $
B_d \to D_{CP}K_S$,
$B_s \to D_{CP}\phi)$ are comparable in
size, and contribute dominantly to the $D^0$ or $\bar D^0$ parts of
$D_{CP}$ respectively to extract the quantity
\beq \label{gamone}
\arg(b \to c \bar u s) + \arg(c \to d \bar d u)
- \arg(b \to u \bar c s) - \arg(\bar c \to \bar d d \bar u)
\equiv \gamma
\eeq
This measurement of $\gamma$ is
manifestly independent of the $B-\bar B$ mixing phase%
\footnote{%
We emphasize that 
$CP$ asymmetries into final states that contain
$D_{CP}$ cannot be affected by possible new contributions to $D - \bar D$
mixing. One
identifies $D_{CP}$ by looking for $CP$ eigenstate decay products like
$K^+K^-$, $\pi\pi$ or $\pi K_S$. 
As $(\Delta \Gamma / \Gamma)_D$ is known to be tiny,
the mass eigenstates cannot be identified. 
The relevant quantity that enters in the
calculation of the $CP$ asymmetry is 
the $D$ meson decay
amplitude and not the $D-\bar D$ mixing amplitude. Thus, the only new
physics in the $D$ sector
that could affect the standard analysis are new contributions
to the $D$ decay amplitudes.}.

The mode $(c7)~ b \to u \bar u d$ (\eg\ $B_d \to \pi \pi$, $B_s \to \rho
K_s$) measures the angles $(\beta+\gamma, \beta'+
\gamma$) in the \sm. 
We can combine this measurement, with the phase 
$(\beta, \beta'$) measured in the $(c1)~b \to c \bar c s$ 
mode to get another determination of $\gamma$ that is
independent of the phase in the $B-\bar B$ mixing matrix 
\eg\ comparing
$a_{CP}(t)[B_d \rightarrow \psi K_S]$ to
$a_{CP}(t)[B_d \rightarrow \pi \pi]$ allows us to extract
\beq \label{gamtwo}
\arg(b \to c \bar c d) - \arg(b \to u \bar u d) \equiv \gamma .
\eeq
Since both of the above evaluations of $\gamma$,
Eqs. (\ref{gamone}) and (\ref{gamtwo}) are
manifestly independent of any phases in the neutral meson mixing
matrices, 
the only way they can differ is if there
are new contributions to the $B$ or $D$ meson decay amplitudes.

The remaining charged current decay mode $(c8)~b \to u \bar u s$ 
suffers from large theoretical uncertainty 
since the tree and penguin contributions are similar in magnitude
and we will not study it here.

For the neutral current modes 
we will first assume that the dominant \sm\ contribution is from a
penguin diagram with a top quark in the loop, and discuss corrections
to this later. Since these are loop
mediated processes even in the \sm, $CP$ asymmetries into final states
that can only be produced by flavor changing neutral current vertices
are likely to be fairly sensitive to the possibility of new physics in
the $B$ meson decay amplitudes. The modes $(n3)~b \to s
\bar d d$ and $(n4)~b \to d \bar d d$ however, result in $CP$
eigenstate final states that are the same as 
for the charged current modes
$(c8)~b \to u \bar u s$ and $(c7)~b \to u \bar u d$ respectively.
Hence they cannot be used to study $CP$ violation, but rather act 
as corrections to the charged current modes.

In the \sm\ the 
mode $(n1)~ b \to s \bar s s$, (\eg\
$B_d \to \phi K_S$, $B_s \to \phi \eta')$ measures
the angle $\beta$ or $0$ in $B_d$ and $B_s$ decays. We can once
again try and isolate new physics in the decay amplitudes by comparing
these measurements with the charged current measurements of $\beta$.
Finally, $(n2)~ b \to d \bar s s$, \eg\
($B_d \to K_SK_S$, $B_s \to \phi K_S)$ 
measures the angle $0$ and $\beta$ for \sm\ $B_d$ and $B_s$ decays.

\subsection{Standard Model Pollution}

In all of the preceding discussion, we have considered the idealized
case where only one \sm\ amplitude contributes to a particular decay
process and we worked to first order
in the Wolfenstein approximation.
We would now like to estimate the size of the sub-leading
\sm\ corrections to the above processes, which then allows us to
quantify how large the new physics effects have to be in order for
them to be probed, and what are
the most promising modes to study.
In this sub-section we concentrate on the charged current modes, and
one neutral current mode, $(n2)~b\to d\bar s s$. We reserve the study
of $(n1)~b\to s\bar s s$ to the next subsection. 

There is a \sm\ penguin contribution to $(c1)~b \to c \bar c s$.
However, as is well known, this contribution has the same phase as the
tree level contribution (up to corrections of order $\beta'$) and
hence $\dphi=0$ in Eq.~(\ref{aa}). Thus in the absence of new
contributions to decay amplitudes,
the decay $B_d \to \psi K_S$ 
cleanly measures the phase $\beta+\delta_{m_d}$ 
(where $\delta_{m_d}$
denotes any new contribution to the mixing phase). The mode $(c2)~b
\to c \bar c d$ also has a penguin correction in the \sm. However, 
in this case 
$\phi_{12}={\cal O}(1)$ and we estimate
the correction as \cite{gronau,gronau-london96}
\beq
\dphi_{SM}(b \to c\bar c d) \simeq
\frac{V_{tb}V_{td}^*}{V_{cb}V_{cd}^*}\frac{\alpha_s(m_b)}{12 \pi}
\log(m_b^2/m_t^2)\lsim 0.1 ,
\eeq
where the upper bound is obtained for $|V_{td}| < 0.02$, $m_t=180$ GeV
and $\alpha_s(m_b)=0.2$. The mode $(c3)~ b \to c \bar u d$ does not get
penguin corrections, however there is a doubly Cabbibo suppressed tree
level correction coming from $(c5)~ b \to u\bar c d$. Thus $B_d \to
D_{CP} \rho$ gets a second contribution with different CKM elements. 
While in general $\dphi$ can be a function of hadronic matrix
elements, 
here we expect this dependence to be very weak \cite{koide}.
In the factorization approximation, the matrix elements  
of the leading
and sub-leading amplitude are identical, as are 
the final state rescattering effects.  
Moreover, both these cases get contributions from only
one electroweak diagram,
thus reducing
the possibility of complicated interference
patterns. We then estimate
\beq
\dphi_{SM}(b \to c\bar u d) =
\frac{V_{ub}V_{cd}^{*}}{V_{cb}V_{ud}^{*}} r_{FA}
\le 0.05 .
\eeq
where $r_{FA}$ is the ratio of matrix elements with $r_{FA}=1$ in the
factorization approximation.
We have used $|V_{ub}/V_{cb}| < 0.11$, and used what we believe is a 
reasonable limit for the matrix elements ratio,
$r_{FA}<2$, to obtain the upper bound. 

The technique proposed to extract $\gamma$
using the modes $(c4)~b \to c \bar u s$ and $(c6)~b \to u \bar c s$
is manifestly independent of any ``\sm\ pollution''.
Finally $(c7)~b \to u\bar u d$ suffers from
significant \sm\ penguin pollution, which we estimate as 
\cite{gronau,gronau-london96} 
\beq \label{pen}
\dphi_{SM}(b \to u \bar ud) \simeq
\frac{V_{tb}V_{td}^{*}}{V_{ub}V_{ud}^{*}}\frac{\alpha_s(m_b)}{12 \pi}
\log(m_b^2/m_t^2)\lsim 0.4 ,
\eeq
where the upper bound is for $|V_{td}| < 0.02$, $|V_{ub}|>0.002$,
$m_t=180$ GeV and $\alpha_s(m_b)=0.2$. The effects of the \sm\ penguin can
be removed by an isospin analysis \cite{Gro-Lon}. 
However, this technique would then also rotate away any 
new physics contributions to the glounic penguin operator.

Finally, $(n2)~b \to d \bar s s$ suffers from an 
${\cal O}(30\%)$ correction due to \sm\ 
penguins with up and charm quarks \cite{fleischer}.

In summary, the cleanest modes are $b \to c \bar cs$ 
and $b \to c \bar us$ since they are essentially free
of any sub-leading effects. The mode $b \to c \bar ud$ and 
suffers only small theoretical uncertainty, less than $0.05$.
For $b \to c \bar cd$ the uncertainty is larger, ${\cal O}(0.1)$,
and moreover cannot be estimated reliably since it depends on the
ratio of tree and penguin matrix elements. 
Finally, the $b \to u \bar ud$ and $b\to d\bar ss$
modes suffer from large uncertainties. 


\subsection {${\bf B \to \phi K_S}$}

In this sub-section we would like to carefully analyse the possibility
of using the $CP$ asymmetry in $B\to\phi K_S$ as a probe of new
physcis.\cite{GIW} 
To this end we carry out a rigorous analysis of
the expected size of the \sm\ pollution. Although we expect a
perturbative estimate of the expected size of the pollution along the
lines of those carried out above for the other decay modes to be 
essentially correct, the
importance of this mode in searching for new physics warrants a more
careful treatment. The sensitivity to new physics of the $B\to\phi K$
decay mode stems from the fact that it is a loop-induced process in
the \sm, and hence could receive contributions from virtual new
physics of comparable size to the \sm\ contribution. 

It is 
well known that in the \sm\ the time--dependent $CP$--violating
asymmetry in  $B_d \to \psi K_S$ [\apks] measures \stb, where
$\beta=\arg (-{V_{cd}V_{cb}^*}/{V_{td}V_{tb}^*})$
and $V_{ij}$ denote the CKM matrix elements. 
Moreover, being dominated by the tree--level transition
$b\to c \bar c s$, the decay amplitude of $B_d \to \psi K_S$ 
is unlikely to receive significant corrections from 
new physics.\footnote{
There is, of course, a possible new contribution to 
the $B^0-\bar B^0$ mixing amplitude. This does not 
affect the generality of our arguments or the conclusions \cite{GrWo}.}
Interestingly, within the 
\sm\ the $CP$  asymmetry in $B_d \to \phi K_S$ [\afks] 
also measures \stb\ if,  
as naively expected,
the decay amplitude is dominated by the
short--distance penguin transition $b\to s\bar s s$ \cite{Lon-Pec}. 
Since $B_d \to \phi K_S$ is a loop mediated 
process within the \sm, 
it is not unlikely that new physics could
have a significant effect on it \cite{GrWo}. 
The expected branching ratio and the high identification
efficiency for this decay suggests 
that \afks\ is experimentally accessible at the early stages of 
the asymmetric $B$ factories. 
Thus, the search for a difference between \apks\ and \afks\ is a 
promising way to look for physics beyond the \sm\ 
\cite{GrWo,n-q,ciuchini,london,barbieri}. 

If, indeed, it turns out that \apks\ is not equal to \afks,
it would be extremely important to know how
precise the \sm\ prediction of them being equal is. In particular, one
has to rule out the possibility of unexpected long distance effects
altering the prediction that \afks\ measures \stb\ in the
\sm.

The weak phases of the
transition amplitudes are ruled by products of CKM matrix elements. 
In the $b \to s q \bar q$ case,
relevant to both $B_d \to \psi K_S$ and $B_d \to \phi K_S$,
we denote these by 
$\lambda^{(s)}_q = V_{qb}V_{qs}^*$.
For the purpose of $CP$ violation studies, it is instructive to 
use CKM unitarity and express any 
decay amplitude as a sum of two terms. In particular,
for $b \to s q \bar q$ we eliminate $\lambda^{(s)}_t$ and write
\beq \label{twoterms-s}
A_f =  \lambda^{(s)}_c A_f^{cs} + \lambda^{(s)}_u A_f^{us}.
\eeq
The unitarity and the experimental hierarchy of the CKM matrix
imply $\lambda^{(s)}_t \simeq \lambda^{(s)}_c
\simeq A\lambda^2
+ {\cal O}(\lambda^4)$ and $\lambda^{(s)}_u = A\lambda^4 e^{i\gamma}$,
where $A \approx 0.8$, $\lambda=\sin\theta_c=0.22$ and
$\gamma$ is a phase of order one.
Thus the first and dominant term is real 
(we work in the standard parametrization).
The correction due to the second term, that is complex and  
doubly Cabibbo suppressed, is 
negligibly small unless $A_f^{us} \gg A_f^{cs}$.

The $A_f^{qs}$ amplitudes cannot be calculated exactly since they  
depend on hadronic matrix elements. 
However, in some cases we can reliably estimate their relative sizes.
For $B \to \psi K_S$ the
dominant term includes a tree level diagram while the CKM-suppressed term 
contains only one-loop (penguin) and higher order diagrams.
This leads to  
$A_{\psi K_S}^{cs} \gg A_{\psi K_S}^{us}$, and thus 
insures that \apks\ measures \stb\ in the \sm.
Since both terms for $B \to \phi K_S$ begin at one-loop order one
naively expects  $A_{\phi K_S}^{cs} \sim A_{\phi K_S}^{us}$. In this case 
\afks\ also measures \stb\ in the \sm\ up to corrections of 
${\cal O}(\lambda^2)$.  However, any unexpected enhancement 
of $A_{\phi K_S}^{us}$ would violate this result.
In particular, 
an enhancement of ${\cal O}(\lambda^{-2}) \sim 25$ (analogous
to the $\Delta I = 1/2$ rule in $K$ decays) leads to ${\cal O}(1)$
violations, and subsequently 
to $\apks \ne\afks$ even in the \sm. 

In this sub-section we argue against this possibility,
presenting different arguments that suggest the pollution of 
$A_{\phi K_S}^{us}$ in $B_d \to \phi K_S$ is very small. 
Moreover, we will propose some experimental tests 
that in the near future could provide quantitative bounds on this
pollution. 

The natural tool to describe the $B$ decays of interest is
by means of an effective $b\to s \bar q q$ Hamiltonian.
This can be generally written as 
\beq
{\cal H}_{eff}^{(s)} = \frac{G_F}{\sqrt{2}}\left\{ 
\lambda^{(s)}_t \sum_{k=3..10} C_k(\mu) Q^{s}_k +
\lambda^{(s)}_c \sum_{k=1,2} C_k(\mu) Q^{cs}_k +
\lambda^{(s)}_u \sum_{k=1,2} C_k(\mu) Q^{us}_k 
\right\}\,, 
\label{Heff}
\eeq
where $Q_k^i$ denote the local four fermion operators  and $C_k(\mu)$ 
the corresponding Wilson coefficients, to be 
evaluated at a renormalization scale $\mu\sim {\cal O}(m_b)$.
For our discussion it is useful 
to emphasize the flavor structure of the operators:
$Q^{qs}_{1,2} \sim \bar b s \bar q q$  and $Q^{s}_{3..8} \sim \bar b s
\displaystyle{\sum_{q=u,d,s,c}} \bar q q$, as well as the order of  
magnitude of their Wilson coefficients: 
$C_{1,2}\sim {\cal O}(1)$ and $C_{3..8}\sim {\cal O}(10^{-2})$.
The estimates of the $C_k(\mu)$ beyond the leading logarithmic 
approximation and the definitions of the $Q^i_k$, 
can be found in \cite{Bur-Fl}. 
To an accuracy 
of $O(\lambda^2)$  in the weak phases, ${\cal H}_{eff}^{(s)}$
can be rewritten as 
\beq \label{approxh}
{\cal H}_{eff}^{(s)} = \frac{G_F}{\sqrt{2}}\left\{ 
\lambda^{(s)}_c \left[ \sum_{k=1,2} C_k(\mu) Q^{cs}_k -
\sum_{k=3..10} C_k(\mu) Q^{s}_k \right] +
\lambda^{(s)}_u \sum_{k=1,2} C_k(\mu) Q^{us}_k \right\}~.
\eeq
It is clear that, when sandwiched between the 
$B_d$ initial state and the $\phi K_S$ final state,
the first term corresponds to 
$A_{\phi K_S}^{cs}$ and the second to $A_{\phi K_S}^{us}$ [{\it cf}
Eq. (\ref{twoterms-s})]. The pollution is then generated by 
$Q^{us}_{1,2}$, corresponding to the $b \to s \bar u u$ transition.

Since the matrix elements of the $Q_k^i$ have to be evaluated 
at $\mu \sim {\cal O}(m_b)$, a realistic estimate of their relative 
sizes can be obtained within perturbative QCD.
We recall that  the $|\phi\rangle$ is an almost pure $|\bar s s\rangle$ state. 
The $\omega-\phi$ mixing angle is estimated 
to be below $5\%$ \cite{phimix,pdg}. 
We neglect this small mixing in the following.
Then, the matrix elements of $Q_{1,2}^{us}$ and $Q_{1,2}^{cs}$
evaluated at the leading order (LO) in the factorization
approximation are identically zero. At LO only $Q_{3..8}$,
i.e. the short--distance $b \to s \bar s s$ penguins, have a non 
vanishing matrix element in $B_d \to \phi K_S$. As a consequence, 
the weak phase of the $B_d \to \phi K_S$ decay amplitude is essentially zero.
Nonetheless, given the large Wilson coefficients of $Q^{qs}_{1,2}$, a 
more accurate estimate of their contribution is required.

At next to leading order (NLO), working in a modified factorization 
approximation, one obtains additional contributions from penguin--like matrix
elements of the operators $Q_{2}^{us}$ and $Q_{2}^{cs}$
\cite{fleisch1}. These have been reevaluated recently, and
shown to be important in explaining the CLEO data on charmless
two--body B decays \cite{ciuchini2,fleischer2,lenz}. However, 
even in this case the $b \to s \bar u u$ 
pollution in $B_d \to \phi K_S$ is very small.
The reason is that, in the limit where we can neglect both the 
charm and the up quark masses with respect to $m_b$, the 
matrix elements of $Q^{us}_{1,2}$ and $Q^{cs}_{1,2}$ 
are identical from the point of view of perturbative QCD (up to
corrections of ${\cal O}(m_c/m_b)\sim 0.3$).
However, the overall contribution of the charm operators $Q^{cs}_{1,2}$ 
is enhanced by a factor $\lambda^{-2}$
with respect to the one of $Q^{us}_{1,2}$.
Thus, either if the $B_d \to \phi K_S$ transition is dominated by 
$Q^{s}_{3-10}$ (short--distance penguins) or if it is 
dominated by $Q^{cs}_{1,2}$ (long--distance charming penguins),
the weak phase is vanishingly small.

Of course one could
not exclude a priori a scenario where the contributions of 
$Q^{s}_{3..8}$ and $Q^{cs}_{1,2}$ cancel each other to 
an accuracy of $O(\lambda^2)$. However, this
extremely unlikely possibility would result in an unobservably small  
$BR(B_d \to \phi K_S)$, rendering this entire discussion moot.

As discussed above, any
enhancement of $\langle \phi K_S |Q^{us}_{1,2}  | B_d \rangle$, 
that could spoil the prediction that \afks\ measures \stb\ in the \sm\
should occur at low energies in order not to be compensated
by a corresponding enhancement of 
$\langle \phi K_S | Q^{cs}_{1,2} | B_d \rangle$.
This possibility is not only disfavored  by the OZI rule 
\cite{ozi},\footnote{This non--perturbative prescription has never 
been fully understood in the framework of perturbative QCD, but
can be justified in the framework of the $1/N_c$ expansion, 
and is  known to work well in most cases and
particularly in the vector meson sector \protect\cite{isgur}.}
but is also 
suppressed by the smallness of the energy range where 
the enhancement should occur with respect to the scale of the process. 
We are not aware of any dynamical
mechanism that could favor this scenario. 
Inelastic rescattering effects in $B$ decays due to
Pomeron exchange have been argued not to be negligible
and to violate the factorization limit \cite{donoghue}. However,
even within this context violations of the OZI rule are 
likely to be suppressed \cite{ozi-viol}.

There are experimental tests of our arguments that can be 
achieved in the sector of $b\to d$ transitions. 
These are described 
by an effective Hamiltonian ${\cal H}_{eff}^{(d)}$ completely similar
to the one in Eq. (\ref{Heff}) except for the substitution 
$s\to d$ in the flavor indices of both CKM factors and four--fermion 
operators.
$SU(3)$ flavor symmetry can be used to obtain relation among several
matrix elements. In particular
\beq
\sqrt{2}\;
\langle \phi K_S  |Q^{us}_{1,2}  | B_d \rangle =
\langle \phi\pi^+ |Q^{ud}_{1,2}  | B^+ \rangle +
\langle K^* K^+|Q^{ud}_{1,2}  | B^+ \rangle\,.
\eeq
($SU(3)$ breaking effects, which are typically at
the 30\% level, are neglected here.)
The coefficients of these matrix elements are, 
however, proportional to  different CKM factors.
This is illustrated in Table 1, where we show the relevant $B$ decay
modes along with the Cabibbo factors corresponding to the leading
and sub--leading contributions to the decay amplitudes.
If our arguments hold, one expects  $BR(B_d \to \phi K_S) 
\sim {\cal O}(\lambda^4)$ and $BR(B^+ \to K^*K^+)$,
$BR(B^+ \to \phi \pi^+) \sim {\cal O}(\lambda^6)$. Notice, however,  that
the overall contribution of $Q^{ud}_{1,2}$ in 
$B^+ \to K^* K^+$  and $B^+ \to \phi \pi^+$ is enhanced with
respect to the one of $Q^{us}_{1,2}$ in $B_d \to \phi K_S$
by the corresponding CKM factors: $\lambda_u^{(d)}/\lambda_u^{(s)} 
= {\cal O}(\lambda^{-1})$. 
Thus, if $\langle \phi K_S  |Q^{us}_{1,2}  | B_d \rangle$
is enhanced by ${\cal O}(\lambda^{-2})$ in order
to interfere with the dominant 
${\cal O} (\lambda^2)$ contributions, then  $BR(B^+ \to \phi \pi^+)$ 
and/or $BR(B^+ \to K^*K^+)$ would be dominated by the similarly enhanced 
matrix elements of $Q^{ud}_{1,2}$. 
This would result in an enhancement of 
the naively Cabibbo suppressed modes, 
i.e. we should observe $BR(B^+ \to \phi \pi^+) \sim {\cal O}(\lambda^2)$ 
and/or $BR(B^+ \to K^*K^+) \sim {\cal O}(\lambda^2)$
[while $BR(B_d \to \phi K_S)$ is still $\sim {\cal O}(\lambda^4)$].
Similar arguments hold for the corresponding $B_d$ decay modes,
however in that case the $SU(3)$ relation is not quite as precise.

\begin{table}
\[ \begin{array}{||c||c|c|c||} \hline\hline
\qquad\quad\mbox{Decay\ mode}\qquad\quad & \multicolumn{3}{c||}{
     \mbox{Operators \ and \ CKM\ factors}   }  \\
&\quad \mbox{penguins}\quad &\quad c\mbox{--trees}\quad &
 \quad u\mbox{--trees}\quad  \\ \hline
B_d \to \phi K_S   & Q^s_{3..8}    & Q^{cs}_{1,2}   & Q^{us}_{1,2}  \\ 
                 & \lambda_t^{(s)} \sim \lambda^2 
                 & \lambda_c^{(s)} \sim \lambda^2 
	         & \lambda_u^{(s)} \sim \lambda^4 \\  \hline
B^+ \to \phi \pi^+ \ {\rm and} \ B^+ \to K^*K^+
                 & Q^d_{3..8}    & Q^{cd}_{1,2}   & Q^{ud}_{1,2}  \\
                 & \lambda_t^{(d)} \sim \lambda^3 
                 & \lambda_c^{(d)} \sim \lambda^3 
	         & \lambda_u^{(d)} \sim \lambda^3 \\ \hline\hline
\end{array} \]
\label{su3}
\vspace{0.1in}
\caption
{$SU(3)$ related $B$ decay modes that allow us to quantify the \sm\
pollution in \afks.}
\end{table}

To get a quantitative bound we define the ratios
\beq
R_1 = \frac{BR(B^+ \to \phi \pi^+ )}{BR(B_d \to \phi K_S)}, \qquad
R_2 = \frac{BR(B^+ \to K^*K^+ )}{BR(B_d \to \phi K_S)}\,,
\eeq
such that in the \sm\ the following inequality holds
\beq 
\left| \apks-\afks \right|
< \sqrt{2}\lambda \left( \sqrt{R_1} + \sqrt{R_2} \right) 
[1+R_{SU(3)}] + {\cal O}(\lambda^2)\,,
\label{limit}
\eeq
where $R_{SU(3)}$ represents the $SU(3)$ breaking effects.
While measuring \afks\ it should be possible to 
set limits at least of order one on $R_1$ and $R_2$ and
thus to control by means of Eq. (\ref{limit}) the accuracy to 
which \afks\ measures $\sin 2\beta$ in the \sm. The limits
$\sqrt{R_1}, \sqrt{R_2} \lsim 0.25$ would reduce the theoretical
uncertainty to the $10\%$ level.

It may be possible to confirm that $BR(B^+ \to \phi \pi^+ )$ and
$BR(B^+ \to K^*K^+ )$ are not drastically enhanced based just on the
current CLEO data. The CLEO colloboration already has reported the
bounds
$BR(B^+ \to \phi \pi^+ ) < 0.56\times 10^{-5}$
(Ref. \citenum{cleo23}) 
and $BR(B^+ \to K^*\pi^+ ) < 4.1\times 10^{-5}$ (Ref. \citenum{cleo}). 
Given the similarity of energetic $K$'s and $\pi$'s in the CLEO
environment,  
it is plausible that a bound similar to the latter
can also be derived for the mode
$B^+ \to K^*K^+$. 
Bounds on these branching
ratios of ${\cal O}(10^{-5})$ would clearly imply that the rates are
not ${\cal O}(\lambda^2)$ as they would be if the matrix elements of
$Q^{ud}_{1,2}$ were enhanced by ${\cal O}(\lambda^{-2})$.

The above experimental test can only confirm that \afks\ measures
\stb\ in the \sm. If it turns out that 
$R_1$ or $R_2$ is large, this may be either
due to the failure of our conjectures or due to new physics. 
If, however, $R_1$ and $R_2$ are small, and  
$\apks - \afks$ violates the \sm\ prediction of Eq. (\ref{limit}), this
would be an unambiguous sign of new physics.

Another possible check of our conjecture could be achieved 
through the measurement of the $CP$ asymmetry in 
$B_d \to \eta' K_S$. Recently CLEO has measured a large branching 
ratio for the related decay $B^+ \to \eta' K^+$, suggesting these
processes are penguin dominated and thus that $a_{CP}(\eta' K_S)$ 
also should measure $\sin2\beta$ in the \sm\ \cite{london}. 
Nonetheless, the $|\eta'\rangle$ has a non negligible 
$|\bar u u\rangle$ component that could enhance the 
$b \to u\bar u s$ pollution and the $\eta'$ mass is 
one of the few exception where the OZI rule 
is known to be badly broken. Thus, without fine tuning, a sufficient 
condition to support our claim on \afks\ could be 
obtained by an experimental evidence of
$a_{CP}(\eta' K_S)=a_{CP}(\phi K_S)$. This would imply that the 
$b \to u\bar u s$ pollution is negligible in both cases. 

To summarize, we have argued that the deviation from the prediction that 
\afks\ measures $\sin 2\beta$ in the \sm\ is 
of ${\cal O}(\lambda^2) \sim 5\%$.
Moreover, we have shown how the accuracy of this prediction
can be tested experimentally.
While we concentrated on the time-dependent $CP$ asymmetry 
it is clear that 
our arguments hold also for direct $CP$ violation in charged and neutral
$B \to \phi K$ decays. Namely, that 
in the \sm\ the direct $CP$ asymmetry is ${\cal O}(\lambda^2)$. 
Experimentally, we can hope to get an accuracy for both the time dependent
and the direct $CP$ violation of about $10\%$.
Therefore, any measurable direct $CP$ violation in $B \to \phi K$
or an indication that $\apks \ne \afks$,
combined with experimental evidence that the \sm\ pollution is of 
${\cal O}(\lambda^2)$
will signal physics beyond the \sm.

\subsection{Models}

In this sub-section we discuss three models that could have
experimentally detectable effects on $B$ meson decay amplitudes, and
violate the \sm\ predictions $(i)$ and $(ii)$. 
We also discuss ways to distinguish these models from each other.

{\bf $(a)$ Effective Supersymmetry:} 
This is a supersymmetric extension of the \sm\ that seeks to retain
the naturalness properties of supersymmetric theories, while avoiding
the use of family symmetries or ad-hoc supersymmetry breaking boundary
conditions that are required to solve the flavor problems generic 
to these models \cite{ckn,dine-kagan}. In this
model, the $\tilde t_L$, $\tilde b_L$, $\tilde t_R$ and the gauginos
are light (below 1 TeV), while the rest of the super-partners are
heavy ($\sim 20$ TeV). The bounds on the squark mixing angles in this
model can be found in \cite{ckn2}. Using the formulae in \cite{masiero}
we find that for $\tilde b_L$ and gluino masses in the $100
- 300$ GeV range, this model generates $b \to s q \bar q$ and $b \to d
q \bar q$ transition amplitudes 
via gluonic penguins that could be up to twice as large as the \sm\
gluonic penguins, and with an unknown phase. Thus this model could
result in significant deviations from the predicted patterns of $CP$
violation in the \sm. We estimate these corrections to be
\beqa
\dphi_A(b \to c \bar c s) \lsim 0.1, \qquad &&
\dphi_A(b \to c \bar c d) \lsim 0.2, \qquad 
\dphi_A(b \to u \bar u d) \lsim 0.8, \qquad \nonumber \\
\dphi_A(b \to s \bar s s) \lsim 1, \qquad &&
\dphi_A(b \to d \bar s s) \lsim 1, \qquad 
\eeqa
%

{\bf $(b)$ Models with Enhanced Chromomagnetic Dipole Operators:}
These models have been proposed to explain the discrepancies between
the $B$ semi-leptonic branching ratio, the charm multiplicity in $B$
decays and the \sm\ prediction for these quantities.
These enhanced chromomagnetic dipole operators 
come from gluonic penguins that arise naturally in TeV
scale models of flavor physics \cite{kagan}. 
In order to explain the above discrepancies with the \sm, these models
have amplitudes for $b \to s g$ that are 
about 7 times larger than the
\sm\ amplitude. 
The $b \to s q\bar q$ transition in this model is dominated by the
dipole operator for $b \to sg$ through the chain $b \to s g^{*} \to
sq\bar q$.
This interferes with the \sm\ $b \to s q\bar q$ amplitude. 
For the $B \to X_s \phi$
the net result is 
that the new amplitudes can be up to 
a factor of two larger than the \sm\
penguins and with arbitrary phase \cite{kagan2}. It is thus
plausible that similar enhancements can be present in 
the exclusive
$b \to c \bar c s$ transitions as well.
In addition, $b \to d g$
can be as large as $b \to s g$. However in the \sm\ the $ b
\to d$ penguins are Cabbibo suppressed compared to the $b \to s$
penguins. Thus in this model the corrections to the $b \to d \bar q
q$ modes could be much larger than the corrections to the $b \to s
\bar q q$ modes.
In the explicit models that have been studied, the
relative corrections to the $b \to dg$ \sm\ amplitude are up to 3 times
larger than those to the \sm\ $b\to s g$ amplitude \cite{kagan2}.
We estimate the following corrections to the dominant \sm\ amplitudes 
\beqa
\dphi_B(b \to c \bar c s) \lsim 0.1, \qquad && 
\dphi_B(b \to c \bar c d) \lsim 0.6, \qquad
\dphi_B(b \to u \bar u d) \lsim 1, \nonumber \\
\dphi_B(b \to s \bar s s) \lsim 1, \qquad &&  
\dphi_B(b \to d \bar s s) \lsim 1.
\eeqa
%

{\bf $(c)$ Supersymmetry without R-parity:}
Supersymmetric extensions of the \sm\ usually assume the existence of
a new symmetry called $R$-parity. However, phenomenologically viable
models have been constructed where $R$-parity is not conserved
\cite{rparity}. In the absence of $R$-parity, baryon and lepton number
violating terms are allowed in the superpotential. Here we assume that 
lepton number is conserved in order to avoid bounds from proton decay
and study the effects of possible baryon number violating terms. The
relevant terms in the superpotential are of the form
$\lpp_{ijk} \bar u_i \bar d_j \bar d_k$,
where antisymmetry under $SU(2)$ demands $j \ne k$. The tree-level
decay amplitudes induced by these couplings are
then given by
\beq
A(b \to u_i \bar u_j d_k) \approx {\lpp_{i3l} \lpp_{jkl} 
\over 2 m_{\qt}^2}, \qquad
A(b \to d_i \bar d_j d_k) \approx {\lpp_{l3j} \lpp_{lik} 
\over 2 m_{\qt}^2}.
\eeq
Note that due to the requirement $i \ne k$ in the neutral current
mode, the decay $ b\to s\bar s s$ will not be corrected.
If we use, $m_{\tilde q} \simeq M_W$ for the squark masses, and assume
that there are no significant cancellations between the (possibly several)
terms that contribute to a single decay, then the bounds for
the relevant coupling constants are 
\cite{CRS}
\beq \label{rareBf}
\lpp_{ibs} \lpp_{ids} \lsim 5 \times 10^{-3} , \qquad
\lpp_{ibd} \lpp_{isd} \lsim 4.1 \times 10^{-3}, \qquad
\lpp_{ubs} \lpp_{cds} \lsim 2 \times 10^{-2}.
\eeq
(We have imposed the
last bound in Eq.~(\ref{rareBf}) by demanding that the new contribution
to the $B$ hadronic width be less than the contribution from the \sm\
$b \to c \bar u d$ decay mode).
These lead to the following corrections to the dominant \sm\
amplitudes 
\beqa
\dphi_C(b \to c \bar c s) \lsim 0.1, \qquad &&
\dphi_C(b \to c \bar c d) \lsim 0.6, \nonumber \\
\dphi_C(b \to c \bar u d) \lsim 0.5, \qquad &&  
\dphi_C(b \to d \bar s s) \lsim 1.
\eeqa
%

The observed pattern of $CP$ asymmetries can also
distinguish between different classes of new contributions to the $B$
decay amplitudes. Here we list a few examples:

\noindent $(1)$ In model $(a)$ the maximum allowable 
relative corrections to the $b\to s$
and the $b \to d$ \sm\ amplitudes are similar in size.
While in model $(b)$ the relative corrections to the
$b \to d$ amplitude can be much larger.

\noindent $(2)$ 
In both models $(a)$ and $(b)$, the neutral current decay $b \to s
\bar s s$ can get significant $[{\cal O}(1)]$ corrections. In model
$(c)$ however, this mode is essentially unmodified. 

\noindent $(3)$ 
The fact that the $b \to c \bar u d$ channel can be
significantly affected in model $(c)$
is in contrast with the other two models.
In those models the new decay amplitudes were penguin
induced, and required the up-type quarks in the final state to be a
flavor singlet ($c\bar c$ or $u\bar u$), thus giving no correction to 
the $b \to c \bar u d$ decay.

\subsection{Discussion}

Table~2 summarizes the relevant decay modes with their \sm\ uncertainty,
and the expected deviation from the \sm\ prediction in 
the three models we gave as examples.
New physics can be probed 
by comparing two experiments that measure the same phase $\phi_0$ in
the \sm\ [see Eq. (\ref{aa})].
A signal of new physics will be if these two measurements differ by an
amount greater than
the \sm\ uncertainty (and the experimental sensitivity) \ie 
\beq
|\phi(B \to f_1) - \phi(B \to f_2)| >
\dphi_{SM}(B \to f_1) + \dphi_{SM}(B \to f_2).
\eeq
Where $\phi(B \to f)$ is the angle obtained
from the asymmetry measurement in the $B \to f$ decay.

\begin{table}
  \begin{tabular}{|c|c|c|c|c|c|c|}\hline
\qquad Mode \qquad &  SM angle $(\phi_0)$ & $\dphi_{SM}$ & $ \dphi_A$ & 
$\dphi_B$ & $\dphi_C$ & $BR$\\
\hline 
$b\to c\bar cs$ & $\beta$ & 0 & $ 0.1 $ & $ 0.1 $ & $ 0.1 $ &
 $7 \times 10^{-4}$
\\ 
$b\to c\bar cd$ & $\beta$ & $ 0.1$ & $ 0.2$ &$ 0.6$ &$ 0.6$ & 
 $4 \times 10^{-4}$ \\
$b\to c \bar ud$ & $\beta$ & $ 0.05$ & 0 & 0 & $ 0.5$ &
 $10^{-5}$ \\
$b\to s\bar ss$ & $ \beta $ & 0.04 & $ 1$ & $ 1$ & 0 &
 $10^{-5}$ \\
$b\to u\bar ud$ & $\beta + \gamma$ & $ 0.4$ & $ 0.4$ & 
$ 1$ & 0  &
 $10^{-5}$ \\
$b\to u\bar cs$ &  $\gamma$  & 0 & 0 & 0 & 0  &
 $10^{-6}$ \\
$b \to d \bar s s$ & $ 0 $ & $ 0.3$ & $ 1$ & $ 1$ &
$ 1$  &
 $10^{-6}$ \\
\hline
\end{tabular}
\vskip 12pt 
\caption
{Summary of the useful modes. The ``SM angle'' entry corresponds to
the angle obtained from $B_d$ decays 
assuming one decay amplitude and to first order in the Wolfenstein 
approximation. The angle $\gamma$ in the mode $b\to u\bar cs$ is 
measured 
after combining with the mode $b\to c\bar us$. 
New contributions to the mixing amplitude would shift all the entries
by $\delta_{m_d}$.
$\dphi$ (defined in Eq.~(\ref{aa}))
corresponds to the (absolute value of the) correction to the universality
prediction within each model:
$\dphi_{SM}$ -- \sm, $\dphi_{A}$ -- Effective Supersymmetry, 
$\dphi_{B}$ -- Models with Enhanced Chromomagnetic Dipole Operators and
$\dphi_{C}$ -- Supersymmetry without R-parity. 
$1$ means that the phase can get any value.
The $BR$ is taken from \cite{BrHo} and is an order of magnitude estimate
for one of the exclusive channels that can be used in each inclusive mode.
For the $b \to c \bar u d$ mode 
the $BR$ stands for the product 
$BR(B_d \to \bar D \rho) \times BR(\bar D \to f_{CP})$
where $f_{CP}$ is a $CP$ eigenstate.
}
\label{sumtab2}
\end{table}

The most promising way to look for new physics effects in decay
amplitudes is to compare all
the $B_d$ decay modes that measure $\beta$ in the \sm\ 
(and the $B_s$ decay modes that measure $\beta'$ in the \sm).
The theoretical uncertainties among all the decays considered
are at most ${\cal O}(10\%)$,
and they have relatively large rates.
The best mode is $B_d \to \Psi K_S$ which has a sizeable rate and negligible 
theoretical uncertainty. This mode should be the reference mode to which all
other measurements are compared.
The $b \to c \bar u d$ and $b \to s \bar s s$ modes are also 
theoretically very clean. In addition, the $b \to s \bar s s$ being a
loop-mediated process in the \sm, is particularly sensitive to new
physics effects.
In both cases the conservative
upper bound on the theoretical uncertainty is less than $0.05$,
and can be reduced with more experimental data. 
Moreover, the 
rates for the relevant hadronic states are ${\cal O}(10^{-5})$ 
which is not extremely small.
Thus, the two ``gold plated'' relations are 
\beq
|\phi(B_d \rightarrow \psi K_S) - \phi(B_d \rightarrow \phi K_S)| < 0.05,
\eeq
and
\beq
|\phi(B_d \rightarrow \psi K_S) - \phi(B_d \rightarrow D_{CP} \rho)| < 0.05. 
\eeq
Any deviation from these two relations will be a clear indication for new 
physics in decay amplitudes.

Although not as precise as the
previous predictions, looking for violations of the relation
\beq
|\phi(B_d \rightarrow \psi K_S) -
\phi(B_d \rightarrow D^+D^-)| < 0.1,
\eeq
is another important way to search for new physics in the $B$ decay
amplitudes. The advantage is that the relevant rates are rather large,
$BR(B_d \rightarrow D^+D^-) \approx 4 \times 10^{-4}$. However, the theoretical
uncertainty is large too, and
our estimate of $10\%$ should stand as a central value of it.
As long as we do not know how to calculate hadronic matrix elements
it will be hard to place a conservative upper bound.

New physics can also be discovered by comparing the two ways to
measure $\gamma$ in the \sm, \ie\ from $b\to c\bar cd$ combined with
$b\to u\bar ud$, and $b\to c\bar us$ combined with $b\to u\bar cs$.
This is not so promising since the
rates are relatively small, and the theoretical uncertainties
are larger. Thus one would require larger effects in order for them to
be observable.
Moreover, isospin
analysis that would substantially reduce the \sm\ uncertainty 
in the $b\to u \bar ud$ would simultaneously remove the 
isospin invariant new 
physics effects from this mode, thus requiring effects in 
the $b\to c\bar us$ mode 
(which were not found in the three models studied here).

\section{Conclusions}

In this lecture we have studied the possibility of using the 
time dependent $CP$ asymmetries in $B_d$ decays to $CP$ 
eigenstates that will be measured at the asymmetric $B$ factories as a
probe of physics beyond the \sm. The types of new physics that could
affect these experiments can be logically divided into two classes:
new $\Delta B=2$ phyiscs, affecting the \bbbar\ mixing amplitude,
and new $\Delta B=1$ physics, affecting the $B$ decay amplitudes. 

We argued that even in the presence of new $\Delta B=2$ physics, we
can use the $CP$ asymmetries in $B\to \psi K_S$ (\apks), and in $B\to
\pi\pi$ (\app) to reconstruct the unitarity triangle in a model
independent way. In practise however, hadronic uncertainties and
discrete ambiguities in the angles of the unitarity triangle 
make this a difficult program to carry out.  
In certain classes of models, such as most models of low energy
supersymmetry, the \kpnn\ decay rates
are not affected by new physics. One could then use these rates to
accurately constrain the unitarity triangle. Moreover, discrete
ambiguities can be removed by a rough measurement of $CP$ asymmetry in
modes such as $B_d\to \rho\pi$. 

We presented a detailed, model independent study of the possibility of
detecting new $\Delta B=1$ physics. This possibility affects the
precisely known patterns of $CP$ violation predicted in the \sm. Thus,
the experiments are potentially sensitive to small effects. 
We pointed out that the $CP$ asymmetry in the rare
$B\to\phi K_S$ decay is particularly sensitive to new physics since it
is a loop-mediated process in the \sm\ that is theoretically clean,
and experimentally accessible.
We undertook a detailed study of the possible 
\sm\ contamination to the sensitivity of this mode and proposed a way
to bound this contamination experimentally. Finally we analysed a
number of models of new physics and showed that not only is it
possible that the $B$ decay amplitudes are modified in an
experimentally discernible way, but that it is possible to
discriminate between classes of models of new physics using these $CP$
violating measurements. 

\medskip

\noindent
{\bf{Acknowledgements}}
I would like to thank Y. Grossman, G. Isidori and Y. Nir for highly
enjoyable colloborations on the topics covered in this lecture.

\end{document}